\documentclass[twocolumn]{aastex631}

\usepackage{graphicx}	
\usepackage{amsmath}	

\usepackage{amssymb}	
\usepackage{bm}
\usepackage{amsfonts}
\usepackage{latexsym}
\usepackage{amsmath}
\usepackage{epsfig}
\usepackage{amsbsy}
\usepackage{enumitem}
\usepackage{multirow}

\begin{document}

\title{Continuous gravitational waves from thermal mountains on accreting
neutron stars: effect of the nuclear pasta phase}

\author[0000-0002-4850-8351]{Hong-Bo Li}
\affiliation{Kavli Institute for Astronomy and Astrophysics, Peking University,
Beijing 100871, China}

\author[0000-0002-1334-8853]{Lijing Shao}
\affiliation{Kavli Institute for Astronomy and Astrophysics, Peking University,
Beijing 100871, China}
\affiliation{National Astronomical Observatories, Chinese Academy of Sciences,
Beijing 100012, China}

\author[0000-0002-3388-1137]{Cheng-Jun Xia}
\affiliation{Center for Gravitation and Cosmology, College of Physical Science
and Technology, Yangzhou University, Yangzhou 225009, China}

\author[0000-0002-9042-3044]{Ren-Xin Xu}
\affiliation{Department of Astronomy, School of Physics, Peking University,
Beijing 100871, China}
\affiliation{Kavli Institute for Astronomy and Astrophysics, Peking University,
Beijing 100871, China}

\begin{abstract}
As density increases, the shape of nuclei transitions to  non-spherical ``nuclear
pasta" structures. The physical properties of the nuclear pasta, such as thermal
conductivity and elasticity, have implications for detecting continuous
gravitational waves from a rapidly rotating neutron star.  In this work, we
investigate the effect of the nuclear pasta on the quadruple moment, and find
out that, compared with previous work, the quadrupole moment contributing to
continuous gravitational-wave radiation can be up to two orders of magnitude
larger. We also discuss the relationship between the quadruple moment and the
maximum shear strain. Considering the properties of nuclear pasta, we study the
detectability of known accreting neutron stars and compare predicted results to
the detectable amplitude limits.  These sources are well above the sensitivity
curves for Cosmic Explorer and Einstein Telescope detectors. Our work advances
the understanding of the properties of nuclear pasta and a possible mechanism
for continuous gravitational waves.
\end{abstract}


\section{Introduction}
\label{sec: intro}

Spinning non-axisymmetric neutron stars (NSs) are promising sources of
continuous gravitational waves (CGWs) via quadrupole radiation \citep[see][for a
detailed review]{Riles:2022wwz, Wette:2023dom, 2024arXiv240302066J}. The
characteristic amplitude of CGW signal scales with the square of the rotation
frequency, combined with the noise curve of ground-based detectors, making the
more rapidly rotating NSs ideal candidates for detection. The detection
strategies for CGWs from rapidly rotating NSs mainly fall into two categories:
targeted observation of accreting millisecond X-ray pulsars
\citep[AMXPs;][]{LIGOScientific:2004sbr, LIGOScientific:2007leh,
LIGOScientific:2011msu, LIGOScientific:2021ozr}; all-sky searches for CGWs from
unobserved sources \citep{LIGOScientific:2007leh, LIGOScientific:2017csd,
LIGOScientific:2019yhl, KAGRA:2021una, KAGRA:2022dwb, Steltner:2023cfk,
Tripathee:2023muh}.  Although CGWs from NSs have yet to be detected, the upper
limit of the amplitude and ellipticity are becoming more strictly limited.

Low-mass X-ray binaries  (LMXBs)  were originally invoked as a strong source of
CGWs to explain an observational puzzle. Observation of NSs in LMXBs has shown
that spins are far below the predicted breakup limit of $1000$\,Hz
\citep{1994ApJ...423L.117C, Lattimer:2006xb}.  For instance, the
fastest-spinning accreting pulsar observed to date is IGR J00291$+$5934 that
spins at $599$\,Hz \citep{Galloway:2005rg}. The statistical evidence about the
spin distributions has been discussed by \citet{Patruno:2017oum}. They found the
distribution to comprise two sub-populations: one at relatively low spin
frequencies with an average of $\approx 300$\,Hz and the other at higher
frequencies with an average of  $\approx 575$\,Hz. In particular, the spin
frequency has a statistical cut-off approaching to $730$\,Hz
\citep{Chakrabarty:2003kt, Patruno:2010qz}. The candidate mechanisms are divided
into two camps. The first is the interaction between the accretion disc and the
magnetosphere of NS \citep{1978ApJ...223L..83G, 1997ApJ...490L..87W,
Andersson:2004zz}. The second possible scenario is that CGW may provide the
torque needed to balance the accretion torque \citep{1978MNRAS.184..501P,
1984ApJ...278..345W, Bildsten:1998ey}.  The latter scenario is our main focus in
the present work.

 Many possible mechanisms may lead to CGW emission. These include: (i) the
 $r$-mode instability \citep{Andersson:1998qs, Levin:1998wa, Heyl:2002pe,
 Mahmoodifar:2013quw, Pi:2014rdd};  (ii) free precession \citep{Jones:2001yg,
 VanDenBroeck:2004wj, Gao:2020zcd, Gao:2022hzd}; (iii) non-axisymmetric
 deformations of the NS crust, including magnetic deformation
 \citep{Melatos:2005ez, Haskell:2007bh, Fujisawa:2022dzp}, elastic deformations
 in the crust \citep{Bildsten:1998ey, Ushomirsky:2000ax, Haskell:2006sv,
 Gittins:2020cvx}, and other forms \citep{Yim:2023nda, Yim:2024eaj}.  The
 associated non-axisymmetric deformations of NSs are commonly referred to as
 ``magnetic mountains'' or ``thermal mountains''.

It is considered that the NS crust exists from the bottom of the ocean layer
with a density in the range of $10^{6} $--$10^{9}\,\rm g \, cm ^{-3}$ inwards to
the boundary with a fluid core at a density of the order of the saturation
density of nuclear matter, $\rho_{\rm s} \approx 2.6 \times 10^{14}\, \rm
g\,cm^{-3}$.  As density increases, the shape of the nuclear matter region
changes from sphere to non-spherical, which is known as ``pasta structure"
\citep[see][for a detailed review]{Caplan:2016uvu, Lopez:2020zne}.  The physical
properties of the nuclear pasta, such as thermal conductivity and elasticity,
have implications for the explanation of cooling curves for MXB 1659$-$29
\citep{Horowitz:2014xca} and of the maximum value for a
spin period of isolated
X-ray pulsars \citep{Pons:2013nea}. Considering the effect of nuclear pasta,
\citet{Sotani:2011nn} calculated the torsional shear mode, which is dependent on
the shear modulus, and explained the quasi-periodic oscillations in SGR
1806$-$20.

For the scenario of the elastic mountains, previous works have ignored the
physical properties of the nuclear pasta and also have not discussed the effect
of the nuclear pasta on the quadrupole moment  \citep{Bildsten:1998ey,
Ushomirsky:2000ax, Haskell:2006sv, Gittins:2020cvx}. Hence, in this paper, we
investigate the physical properties mentioned above, including thermal
conductivity and elasticity.  Considering the effect of the thermal conductivity
for a nuclear pasta, we calculate the temperature asymmetry in the accreted
crust. Next, we calculate the quadruple moment, $Q_{22}$, using the crust
perturbation equations and compare our predicted results to the detectable
amplitude limits. Finally, we discuss the relationship between $Q_{22}$ and the
maximum shear strain.

The paper is organized as follows. In Sec. \ref{sec: Background Structure}, we
compute the hydrostatic structure of an accreting NS for accreted crustal EOS
\citep[as described by][]{1990A&A...229..117H, 1990A&A...227..431H}. The thermal
structure of a spherically symmetric NS is presented in Sec. \ref{sec: TS}.
Based on the background solutions, in Sec. \ref{sec: Temperature perturbation},
we integrate the thermal perturbation equations to determine the magnitude of
the temperature variations.  In Sec. \ref{sec: elastic deformation}, we
calculate the elastic deformation of the accreted crust and compare predicted
results to the detectable amplitudes.  In Sec. \ref{sec: Breaking strain}, we
calculate the maximum values for shear strain and discuss the relationship
between the quadruple moment and maximum shear strain. Finally, we summarize and
discuss our results in Sec.~\ref{sec: summary}.

\section{Structure of an accreting neutron star}
\label{sec: Background Structure}

\begin{figure}
    \centering 
    \includegraphics[width=8cm]{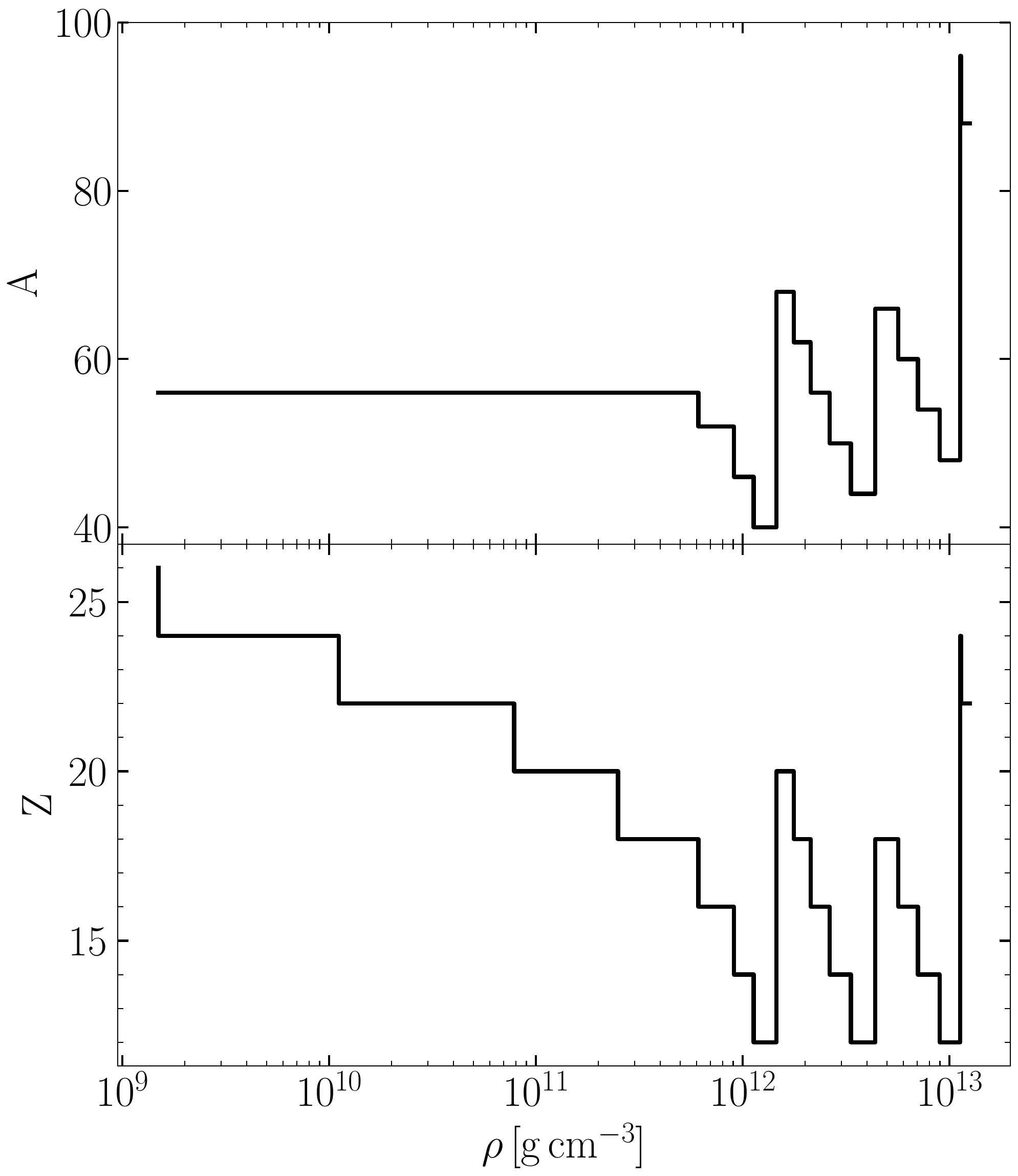}
    \caption{Mass number $A$ (top) and nuclear charge $Z$ (bottom) with
    increasing density. The data are taken from Table 2 of
    \citet{1990A&A...229..117H}. }
    \label{fig: composition} 
\end{figure}

To construct the hydrostatic structure of an accreting NS, we choose the EOS DH
\citep{Douchin:2001sv}, which describes the core,  and use the model of  EOS HZ
\citep{1990A&A...229..117H, 1990A&A...227..431H} in the accreted crust. We
calculate the structure of the crust by employing Newtonian gravity.
\citet{1990A&A...229..117H} have studied the evolution of composition in the
accreted crust and provided tabulated EOS data in detail. For illustration,  we
plot the mass number $A$ and nuclear charge $Z$ as a function of the density in
Fig.~\ref{fig: composition}.

Strictly speaking, the capture layers have a finite thickness. The thickness of
the capture layers was calculated by \citet{Ushomirsky:2000ax}. However, we
follow the approach of  \citet{Osborne:2019iph} and \citet{Hutchins:2022chj},
who assume that the transitions between capture layers are infinitely sharp and
smear the heat released at each transition over shells of constant $(A, Z)$.
This approximation was used to model crustal heating of an accreting NS
\citep{Brown:1999dk,  Brown:2009kw}.

Following \citet{Osborne:2019iph}, we calculate the mass and radius of the core
using the TOV equations and obtain the mass and thickness of the accreted crust
employing Newtonian theory.  The transition density of the interface between the
core and crust is $\rho=1.25 \times 10^{13}\, \rm erg \, cm^{-3}$, and the
transition density of the interface between the crust and ocean is  $\rho=1.49
\times 10^{9}\, \rm erg \, cm^{-3}$. 

\begin{figure}
    \centering 
    \includegraphics[width=8.5cm]{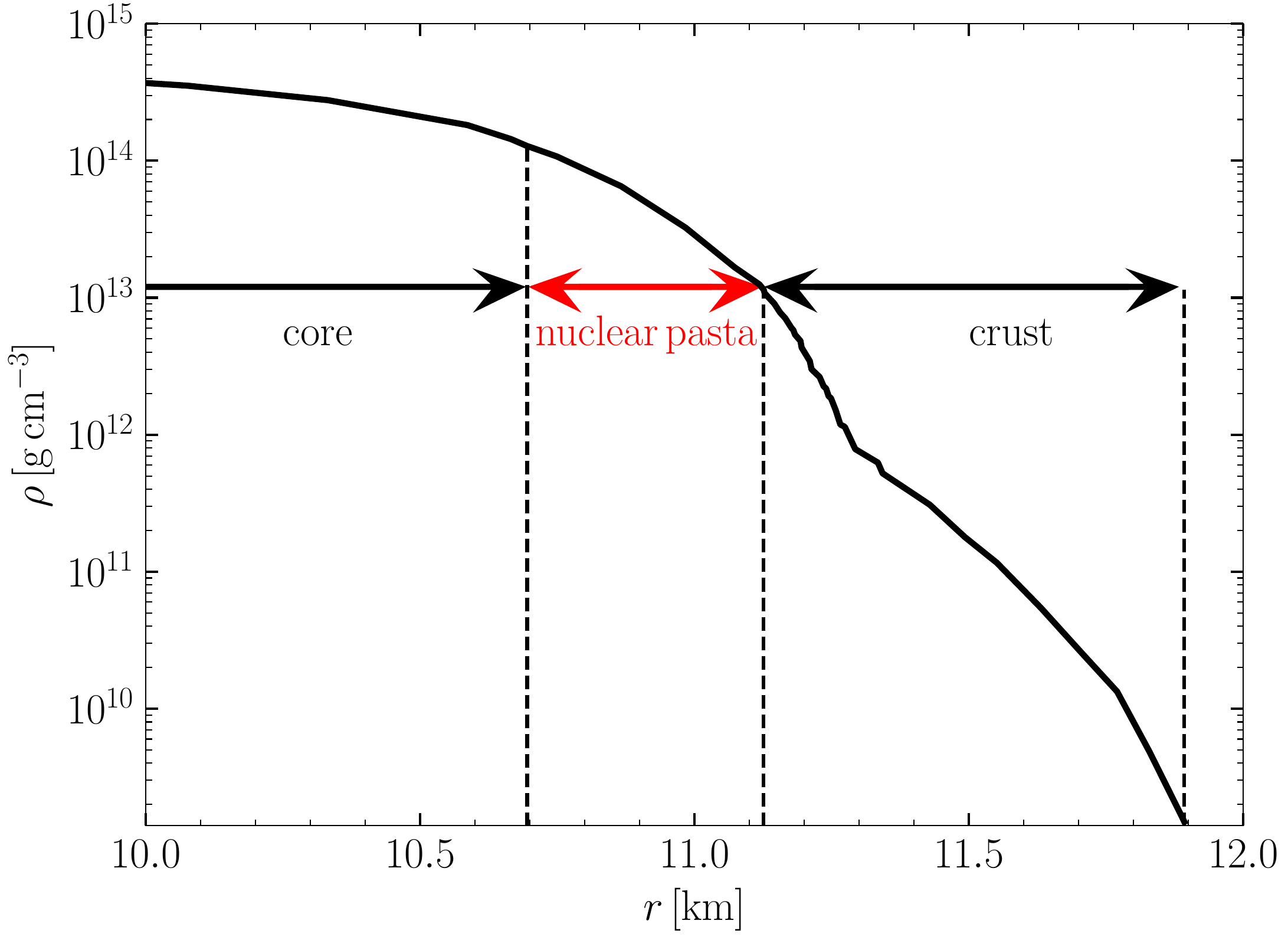}
    \caption{Density as a function of radius, with a central density of
    $\rho_{c}=1.496 \times 10^{15} \rm g\,cm^{-3}$. This model represents a NS
    with mass of $1.84 \, M_{\odot}$ and includes a nuclear pasta thickness of
    $0.43\, \rm km$ and a crust thickness of $0.77\, \rm km$. }
    \label{fig: structure_pasta} 
\end{figure}

In the present work,  we mainly consider the physical properties of the nuclear
pasta so that the transition density of the interface between the core and
nuclear pasta is $\rho=1.28 \times 10^{14}\, \rm erg \, cm^{-3}$. The normal
crust density ranges  from $1.25 \times 10^{13}\, \rm erg \, cm^{-3}$ to $ 1.49
\times 10^{9}\, \rm erg \, cm^{-3}$.  In the following thermal calculations, we
will restrict our computational domain to a range of densities
\citep{1990A&A...229..117H}, appending the density ranges of the nuclear pasta.
Similarly, we calculate the mass and radius of the nuclear pasta and accreted
crust using Newtonian theory.  Hence, the density and radius at the bottom of
the nuclear pasta are
\begin{align}
\rho_{\rm I} &=1.28 \times 10^{14}\, \rm erg \, cm^{-3} \,, \\
R_{\rm I} &= 10.69\, \rm km \,,
\end{align}
and at the interface between the nuclear pasta and normal crust are 
\begin{align}
\rho_{\rm II} &=1.25 \times 10^{13}\, \rm erg \, cm^{-3} \,, \\
R_{\rm II} &= 11.12\, \rm km \,,
\end{align}
and at the interface between the crust and the ocean are
\begin{align}
\rho_{\rm III} &=1.49 \times 10^{9}\, \rm erg \, cm^{-3} \,, \\
R_{\rm III} &= 11.89\, \rm km \,.
\end{align}
The density profile with central density $\rho_{c}=1.496 \times 10^{15}\,  \rm
g\,cm^{-3}$ for our model is shown in Fig. \ref{fig: structure_pasta}. This
model has a mass of $1.84 \, M_{\odot}$ and a nuclear pasta thickness of $0.43\,
\rm km$ and a crust thickness of $0.77\, \rm km$.

Because the EOS of NS is usually weakly dependent on temperature, the
calculation of the thermal structure is decoupled from the hydrostatic
calculation. The crust temperature is proportional to the accretion rates. For a
higher accretion rate, the thermal energy is larger than the Coulomb energy and
leads to the melting of the accreted crust  \citep{Brown:1999dk,
Ushomirsky:2000ax}.  The ratio of Coulomb energy to thermal energy is defined by
\begin{equation} \label{eq: Gamma}
	\Gamma \equiv \frac{(Ze)^2}{k_{\rm B}T}  \left(\frac{4\pi n_{\rm
	ion}}{3}\right)^{1/3}  \,, 
\end{equation}
where  $n_{\rm ion}$ is the ion number density, and  $k_{\rm B}$ is the
Boltzmann's constant. The crystallization point occurs at $\Gamma = 175$
\citep{Haensel:2007yy}. We assume that the ratio value exceeds 175, that is, the
crust and pasta phase are solid in our model.

\section{Thermal structure of an accreting neutron star}
\label{sec: TS}

In this section, we calculate a steady-state thermal structure based on the
Newtonian formulation \citep{Ushomirsky:2000ax, Osborne:2019iph, Hutchins:2022chj}.  The temperature profile is
spherically symmetric and independent of the time.  From energy conservation,
the heat flux $\bf F$ is related to the net rate of heat energy generation per
unit volume $ Q$ via
\begin{equation} \nabla \cdot {\bf F} = Q = Q_{\rm nuc} - Q_{\rm neu} \,,
\end{equation}
where $Q_{\rm nuc}$ is the local energy deposited via nuclear reactions, and
$Q_{\rm neu}$ is the local energy loss due to neutrino emission. The heat flux
is related to the temperature $T$ by the Fourier's law,
\begin{equation} {\bf F} = - \kappa \nabla T \,, \end{equation}
where $\kappa$ is the thermal conductivity.  Assuming the background structure
is spherically symmetric, the heat flux only depends on the radial coordinate
$r$. We can obtain the following system of equations for the heat flux $F$ and
temperature $T$,
\begin{align} \frac{{\rm d} F}{{\rm d} r} & = Q - \frac{2}{r} F  \label{eq:
dFdr} \,, \\ \frac{{\rm d} T}{{\rm d} r} & = - \frac{1}{\kappa}  F   \label{eq:
dTdr}\,.  \end{align}
To solve the above equations, we will consider the description of the heating
term $Q_{\rm nuc}$, neutrino cooling $Q_{\rm neu}$, thermal conductivity, and
the boundary conditions at each end of the integration.  These aspects will be
discussed in detail in the following subsections.

\subsection{Accretion heating}
\label{subsection: Accretion_heating}

In the regions of the crust, we use the EOS HZ model to describe an accreted crust.  The values of mass number A,
nuclear charge Z, electron chemical potential $\mu_{e}$, and the mass fraction
of free neutrons $X_{n}$ are listed in Table 2 of  \citet{1990A&A...229..117H}.
This is important as the nuclear composition changes in discrete steps with
increasing density, corresponding to different electron capture layers. 

As the material moves deeper into the crust, electron capture reactions begin to
occur at a constant pressure, depositing heat into the crust
\citep{1990A&A...227..431H}. The heat deposited in each electron capture layer
per unit volume per unit time depends on the accretion rate $\dot{M}$.  The heat
released in each capture layer is by smearing the heat deposited over whole
shells as \citep{Hutchins:2022chj},
\begin{equation} \label{eq: nuc}
     Q^{\rm crust}_{\rm nuc} =
     \frac{\dot{M}\epsilon_{\rm{nuc}}}{\frac{4}{3}\pi(r^3_{i}-r^3_{i+1})} \,,
\end{equation}
where $\epsilon_{\rm nuc}$ is the heat deposited per nucleon by the relevant
nuclear reaction, and $r_{i}$  is the radii at the $i^{\rm{th}}$ capture layer. 


In the regions of nuclear pasta, we adopt the model of EOS DH, which describes a
non-accreted NS.  The heat deposited per nucleon is $\epsilon_{\rm nuc} =0 $,
meaning $Q^{\rm pasta}_{\rm nuc}=0 $. In our calculation, we only use the value
of pressure $P$ and density $\rho$ to obtain the thickness of the nuclear pasta.

\subsection{Neutrino cooling}
\label{section: Neutrino_cooling}


In the regions of the crust, the neutrino luminosity is dominated by the
electron-ion bremsstrahlung \citep{Brown:1999dk}.  The local energy loss due to
neutrino emission $Q^{\rm crust}_{\rm neu}$ is given by \citep{Haensel:1996rd} 
\begin{equation} \label{eq: neu_1}
     Q^{\rm crust}_{\rm neu} = 3.229 \times 10^{17} \rho_{12}
     T^{6}_{9}\frac{Z^2}{A}(1 - X_{n}) \, \rm erg \, s^{-1}\,cm^{-3} \,,
\end{equation}
where $\rho_{12} = \rho/10^{12}\, \rm g\,cm^{-3}$ and $T_{9} = T/10^9\, \rm K$.

\citet{Lin:2020nxy} have discussed the fast neutrino cooling of nuclear pasta, using the molecular dynamic simulations. 
The value of 
the neutrino luminosity is $Q^{\rm pasta}_{\rm neu}
\approx 8 \times 10^{21} T^{8}_{9}\,\rm erg \, s^{-1}\,cm^{-3}$.
However, in the present work, we assume that the electron-ion bremsstrahlung is still dominant but independent on A, Z, and $X_{n}$. The local
energy loss due to neutrino emission, $Q^{\rm pasta}_{\rm neu}$, can be
approximated by
\begin{equation} \label{eq: neu_2}
     Q^{\rm pasta}_{\rm neu} \approx  3 \times 10^{17} \rho_{12} T^{6}_{9} \,
      \rm erg \, s^{-1}\,cm^{-3} \,.
\end{equation}

\subsection{Thermal conductivity}
\label{subsection: Thermal_conductivity}

The thermal conductivity in the crust helps determine the temperature profile
and heat flux, which is very important for the cooling and transport properties
of the NS \citep{Potekhin:2015qsa}. Using the results of
\citet{Yakovlev:2000jp}, the conductivity can be written as
\begin{equation} \label{eq: conductivity_1}
    \kappa^{\rm crust} = \frac{\pi^2}{3} \frac{k_{B}^2 T n_{e}}{m^{*}_{e}} \tau
    \,.
\end{equation}
Here $n_{e}$ is the electron number density, $m^{*}_{e}$ is the effective
electron mass, and $\tau = {1}/{\nu}$ is the relaxation time with $\nu$ the
scattering frequency of the different  mechanisms.  In the regions of the crust,
the scattering frequency $\nu$ is determined as 
\begin{equation} \label{eq: tau_1}
   \nu = \nu_{\rm ep} + \nu_{\rm eQ} \,,
\end{equation}
where $\nu_{\rm ep}$ and $\nu_{\rm eQ}$ are the scattering frequencies for the
electron-phonon and electron-impurity, respectively.  We neglect the term of the
electron-electron scattering frequency because the strong degeneracy of the
electrons restricts the available phase space \citep{Brown:1999dk}.

 We follow the formalism of \citet{1980SvA....24..303Y}, and the electron-phonon
 scattering is temperature dependent as
\begin{equation} \label{eq: tau_2}
  \nu_{\rm ep} = \frac{12 e^2 k_{B} T}{\hbar^2 c} \,.
\end{equation}
As the density increases,  electron-impurity scattering becomes more important,
and the scattering frequency is written as
\begin{equation} \label{eq: tau_3}
  \nu_{\rm eQ} = \frac{4\pi Q_{\rm imp} e^4 n_{\rm ion}}{p^2_{F} v_{F}}
  \Lambda_{\rm imp} \,,
\end{equation}
where $p_{F}$ and $v_{F}$ are the momentum and velocity of electrons at the
Fermi surface and $\Lambda_{\rm imp} = 1$ is the logarithmic Coulomb factor, and
the impurity parameter is defined by 
\begin{equation}
 Q_{\rm imp}  \equiv  n_{\rm ion}^{-1} \sum_{i} n_{i} (Z_{i} - \langle Z \rangle
 )^2 \,.
\end{equation}
Here, the sum $i$ is over all the different species of ions, with atomic number
$Z_{i}$ and mean atomic number $\langle Z \rangle$. The impurity parameter in
the accreted crust remains highly uncertain. Observational results of the
cooling NS transient MXB 1659$-$29 indicate that the impurity parameter was
constrained to $Q_{\rm imp} < 10 $ \citep{Brown:2009kw}. However,
\citet{Ootes:2019uvd} investigated the long-term temperature evolution of a
transiently accreting NS and suggested that the impurity parameter can range
from $30$ to $100$.

We next briefly review the thermal conductivity of nuclear pasta.
\citet{Horowitz:2008vf} investigated the thermal conductivity of nuclear pasta
using molecular dynamics simulations, and they found that the thermal
conductivity lies in the range of $10^{20} \sim 10^{21}\, \rm erg \, K^{-1}\,
cm^{-1}\, s^{-1}$. Note that this value is from electrons. Topological defects
in nuclear pasta can reduce the thermal conductivity by up to an order of
magnitude \citep{Schneider:2016zyx}.  \citet{Deibel:2016vbc} studied the
observed late-time cooling of MXB 1659$-$29 and pointed out that it is
consistent with low thermal conductivity for nuclear pasta, with a value of
$10^{17}\, \rm erg \, K^{-1}\, cm^{-1}\, s^{-1}$. The thermal conductivity of
nuclear pasta is expected to be sensitive to the temperature and fraction of
species \citep[see][and references therein]{Nandi:2017aqq, Dorso:2020zhk,
Caplan:2020ewl}. In this work, we adopt the following formalism of the thermal
conductivity for nuclear pasta,
\begin{equation} \label{eq: conductivity_2}
    \kappa^{\rm pasta} \approx 3 \times 10^{19}  \left( \frac{T}{T_{8}}\right)\,
    \rm erg \, K^{-1}\, cm^{-1}\, s^{-1} \,.
\end{equation}
The effect of the thermal conductivity of nuclear pasta on the temperature
profiles and heat flux is discussed in Sec.~\ref{Results: BTS}.

\subsection{Boundary conditions }
\label{sec: BC_1}

In this subsection, we describe the boundary conditions to solve  Eqs.~(\ref{eq:
dFdr}) and (\ref{eq: dTdr}).  The bottom boundary is the interface between core
and crust or nuclear pasta.  We assume that the core is isothermal, maintaining
the same temperature as the bottom boundary.  The bottom boundary can be written
as 
\begin{equation}
 F_{\rm bottom} = - \frac{L_{\rm core}}{4 \pi R_{\rm bottom}} \,,
\end{equation}
where $R_{\rm bottom}$ is the raduis of NS core. The neutrino luminosity of the
core due to the modified Urca formula is \citep{Shapiro:1983du}
\begin{align}
L_{\rm core} &  = 5.93 \times 10^{39} \, \rm erg \,s^{-1}\,
\left(\frac{M}{M_{\odot}} \right) \left(\frac{\rho_{\rm nuc}}{\rho}
\right)^{1/3} \nonumber
\\ & \, T^8_8 \exp\left(-\frac{\Delta}{k_{\rm B} T_{\rm bottom}} \right) \,,
\end{align}
where $\rho_{\rm nuc}$ is nuclear density. In this work, we only consider the
normal core, for which $\Delta = 0$.  The case of the superfluid core will be
addressed in future calculations.

The top boundary is set to the steady burning temperature of the hydrogen/helium
burning layer and the accretion as
\begin{equation}
 T_{\rm top}\approx 5.3 \times 10^8 \, {\rm K}\,
 \left(\frac{\dot{m}}{\dot{m}_{\rm Edd}} \right)^{2/7} \,,
\end{equation}
where $\dot{m}$ is the local accretion rate \citep{Schatz:1999kx} and
$\dot{m}_{\rm Edd}$ is the local Eddington limit. In the following calculations,
we assume the uniform accretion over the surface of NS and parameterize our
results in terms of the global accretion rate, $\dot{M}$. 

\subsection{Results: Background thermal structure}
\label{Results: BTS}

\begin{figure}
    \centering 
    \includegraphics[width=8cm]{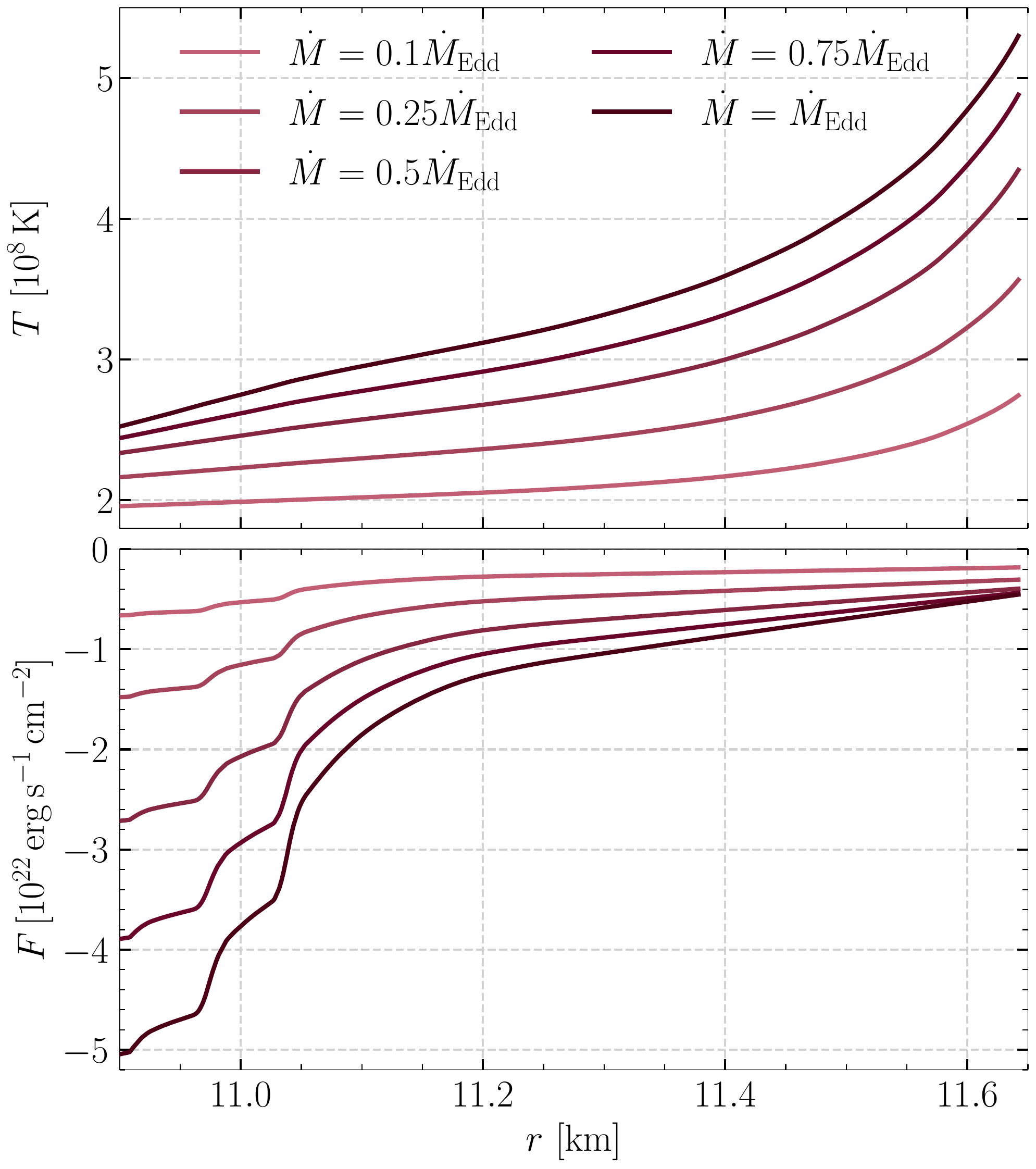}
    \caption{The upper panel shows the temperature profiles against the radius
    without the pasta phase. Different accretion rates are chosen. The lower
    panel shows the heat flux  versus the radius. We set the impurity parameter
    $Q_{\rm imp} =1 $. }
\label{fig: BS_1}
\end{figure}

In the upper panel of Fig.~\ref{fig: BS_1}, we show the temperature
distributions without nuclear pasta. As the accretion rate increases, the
temperature variation becomes more significant. The temperature gradient is
positive, indicating an inwards flow of heat throughout the entire crust.  In
the lower panel of Fig.~\ref{fig: BS_1}, we show the heat flux against the
radius. One notices that the heat flux is negative, which means that heat is
flowing toward the core.

\begin{figure}
    \centering 
    \includegraphics[width=8cm]{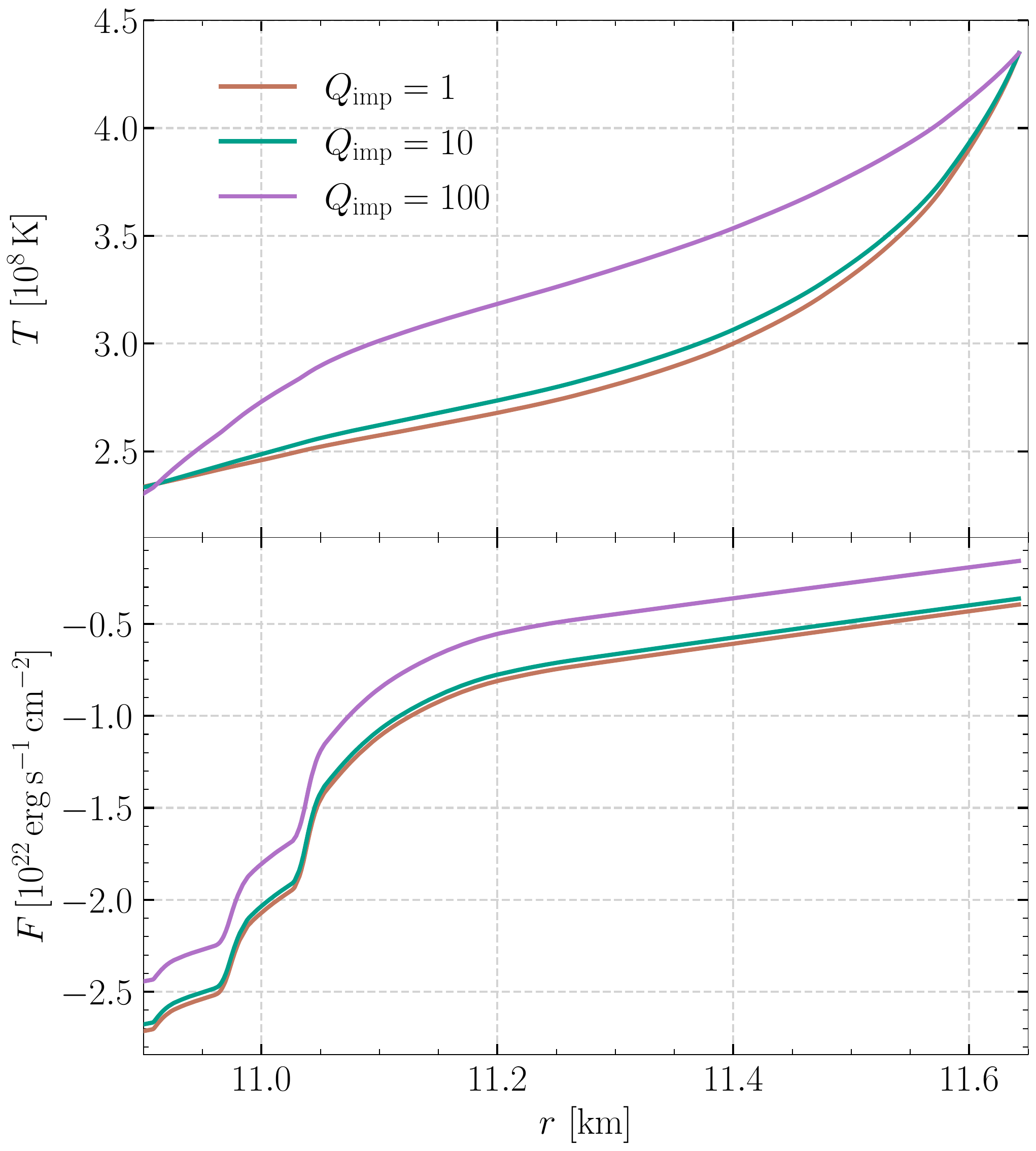}
    \caption{Temperature profiles ({\it upper}) and heat flux ({\it lower}) for
    different impurity parameters without nuclear pasta. The accretion rate
    $\dot{M} = 0.5 \dot{M}_{\rm Edd} $.}
    \label{fig: BS_2}
\end{figure}

Based on the observational constraints, we set the value of the impurity
parameter $Q_{\rm imp} =1 $ to calculate the temperature profiles and heat flux.
However, as mentioned above, the impurity parameter is highly uncertain.  For
example, $Q_{\rm imp} \sim 100 $  is typical of the mixture from the ashes of
steady state burning \citep{Schatz:1999kx}. 

In Fig.~\ref{fig: BS_2}, we show the temperature profiles and heat flux for
different values of $Q_{\rm imp} $. When  $Q_{\rm imp} \lesssim 10 $, these
results are insensitive to the impurity parameter, when $Q_{\rm imp}  \sim 100
$, large differences appear in the profiles of the temperature and heat flux,
because $Q_{\rm imp}$ quantifies the thermal conductivity of the crust, which is
dominated by electron-impurity scattering. The thermal conductivity in the crust
decreases with increasing impurity parameters. This reduces heat conduction from
the crust to the core, resulting in a higher crust temperature and increasing
cooling through crustal neutrino emission.
 
Considering various accretion rates and impurity parameters, we plot the
temperature profiles with nuclear pasta against the radius in Figs.~\ref{fig:
BS_3} and~\ref{fig: BS_4}. We find that the temperature decreases with depth in
the crust, extending the region of the nuclear pasta. The temperature profile is
insensitive to the impurity parameter in the region of the nuclear pasta.
 
\begin{figure}
    \centering 
    \includegraphics[width=8cm]{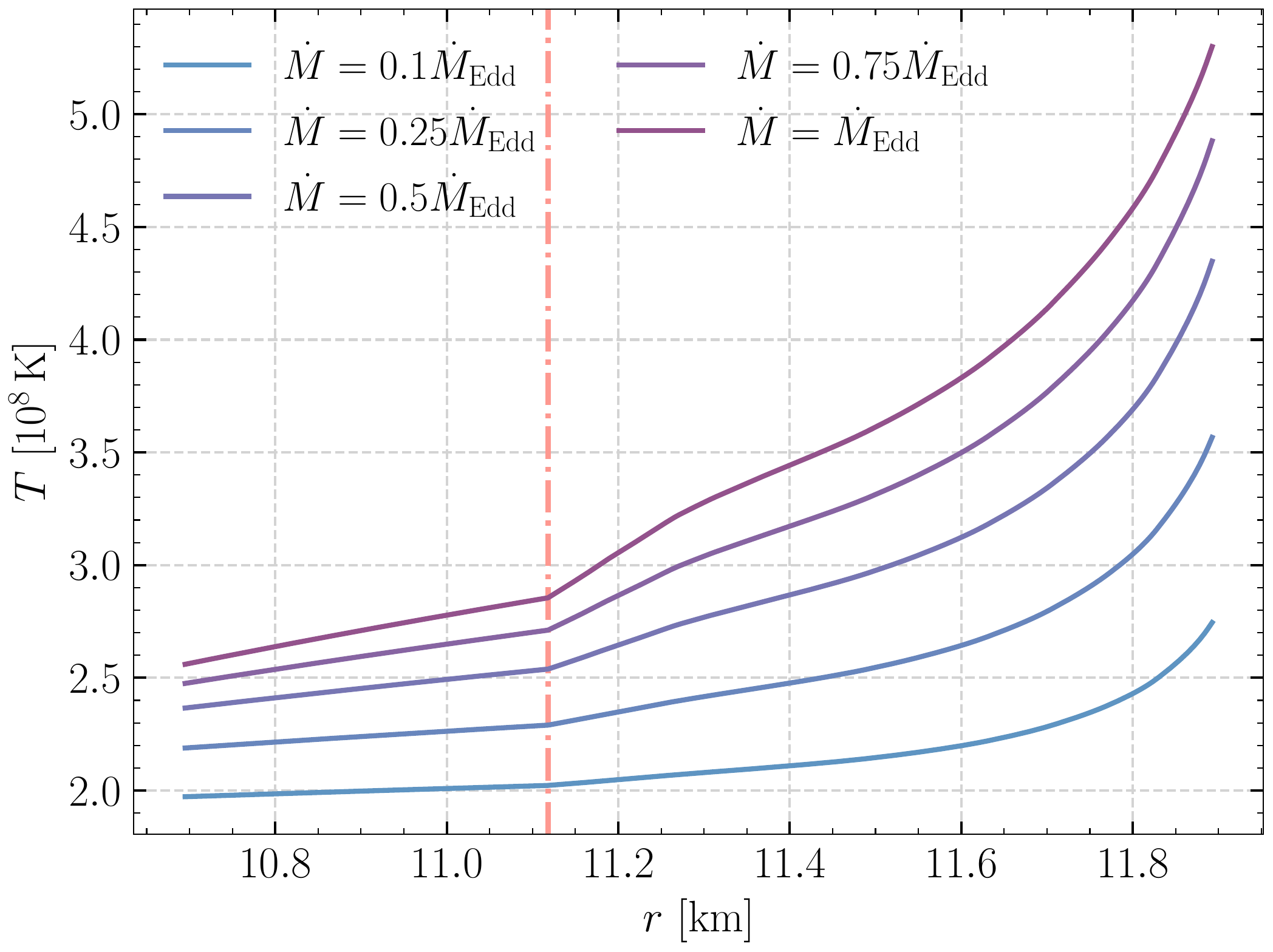}
    \caption{Temperature profiles versus the radius with pasta phase for
    different accretion rates.  The impurity parameter is $Q_{\rm imp} =1 $. The
    vertical dash-dotted line is the nuclear pasta-crust interface. }
    \label{fig: BS_3}
\end{figure}

\begin{figure}
    \centering 
    \includegraphics[width=8cm]{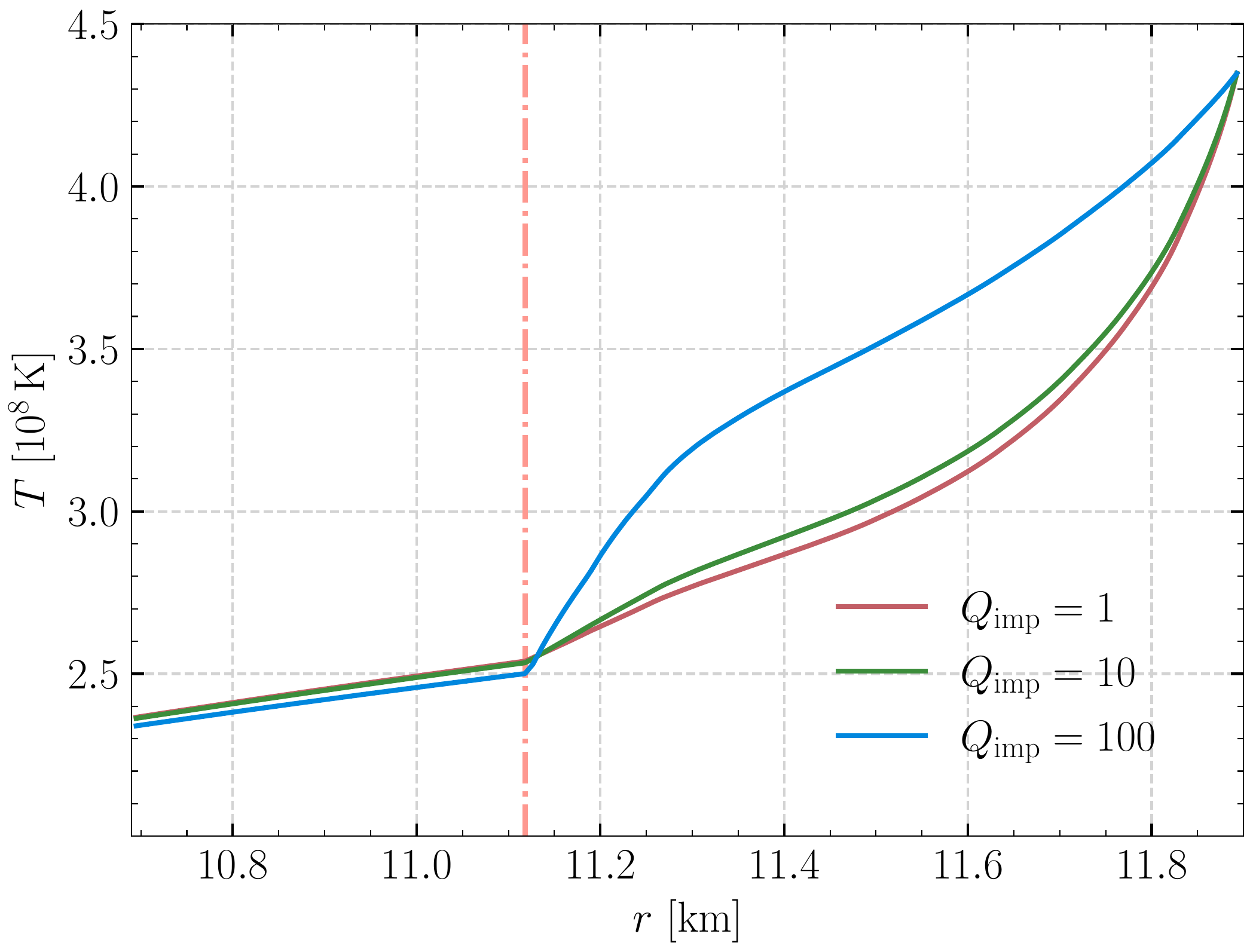}
    \caption{Temperature profiles for different impurity parameters with nuclear
    pasta. The accretion rate is  $\dot{M} = 0.5 \dot{M}_{\rm Edd} $. The
    vertical dash-dotted line has the same meaning as in Fig.~\ref{fig: BS_3}. }
    \label{fig: BS_4}
\end{figure}

\section{Temperature perturbation due to physical asymmetries}
\label{sec: Temperature perturbation}

To calculate lateral temperature variation, $\delta T$, we consider the
following two possible causes. First, the heat deposited, $\epsilon_{\rm nuc}$,
has lateral variation owing to the nuclear transmutation of different elements
depositing different amounts of energy. The lateral variations of this type are
represented by $f_{\rm nuc} = \delta \epsilon_{\rm nuc} / \epsilon_{\rm nuc}$.
Second, the charge-to-mass ratio, $Z^2/A$, may differ if the process of nuclear
burning is different on different sides of the star. The crust conductivity
scales as $\kappa^{\rm crust}\propto (Z^2/A)^{-1}\rho T^{n_{\rm k}}$, where
$n_{\rm k} = 1$ \citep{Schatz:1999kx, Ushomirsky:2000ax}, and neutrino emission
scales as $Q^{\rm crust}_{\rm neu} \propto(Z^2/A)\rho T^{n_{\rm e}}$, where
$n_{\rm e} = 6$ (see Eq.~\ref{eq: neu_1} in this paper). Using the results of
\citet{Ushomirsky:2000ax}, we define the perturbation of charge to mass ratio as
$f_{\rm comp} = \delta (Z^2/A) /  (Z^2/A) $. The  calculation of $\delta T$
depends on $f_{\rm nuc}$ and $f_{\rm comp}$ in the accreted crust.  In the
nuclear pasta regions, $f_{\rm nuc} = f_{\rm comp} = 0$, but the neutrino
emission energy $Q^{\rm pasta}_{\rm neu}$ and thermal conductivity $\kappa^{\rm
pasta}$ can affect the lateral temperature variation $\delta T$. Detailed
results are shown in Sec.~\ref{sec: PTS_Results}.

\subsection{The thermal perturbation equations}

We assume that the angular dependence is given by the spherical harmonic
function $Y_{\ell m}$. The temperature perturbative variable can be written as
$\delta T (r, \theta, \phi)=\delta T(r)Y_{\ell m}$. We can obtain the following
radial perturbation equation~\citep[see][for a detailed variational
derivation]{Ushomirsky:2000ax},
\begin{eqnarray}\label{eq: delta_T}
&&\frac{1}{r^2}\frac{\rm d}{{\rm d} r} \left (r^2 \kappa \frac{{\rm d} \delta
T}{{\rm d} r}\right) -n_{\rm k} \frac{F}{\kappa T} \kappa \frac{{\rm d} \delta
T}{{\rm d} r} - \ell (\ell+1)\frac{\kappa \delta T}{r^2} \nonumber \\
&&-\left\{n_{\rm k} \frac{F^2}{\kappa T}  + \rho \Big[ n_{\rm k} (Q_{\rm nuc}
-Q_{\rm neu})+n_{\rm e} Q_{\rm neu} \Big] \right\}\frac{\delta T}{T} \nonumber\\
=&&\rho\bigg\{f_{\rm comp}( 2 Q_{\rm neu} - Q_{\rm nuc}) - f_{\rm nuc}Q_{\rm
nuc} \bigg\} \,.
\end{eqnarray}
For the case without nuclear pasta, we use the same boundary conditions as in
\citet{Ushomirsky:2000ax}. The temperature perturbation vanishes at the top and
the bottom of the crust. For the case of nuclear pasta, the temperature
perturbation $\delta T = 0 $ extends to the bottom of the nuclear pasta.  

An important issue to clarify is the role of the core. The thermal conductivity
of the core is higher than that of the crust of the NS \citep{Baiko:2001cj}, and
the temperature perturbation in the core is significantly smaller than in the
crust and pasta phase. Therefore, we can approximate the core as perfectly
conducting and isothermal. 

\subsection{Results: perturbed thermal structure}
\label{sec: PTS_Results}

\begin{figure*}
    \centering 
    \includegraphics[width=12cm]{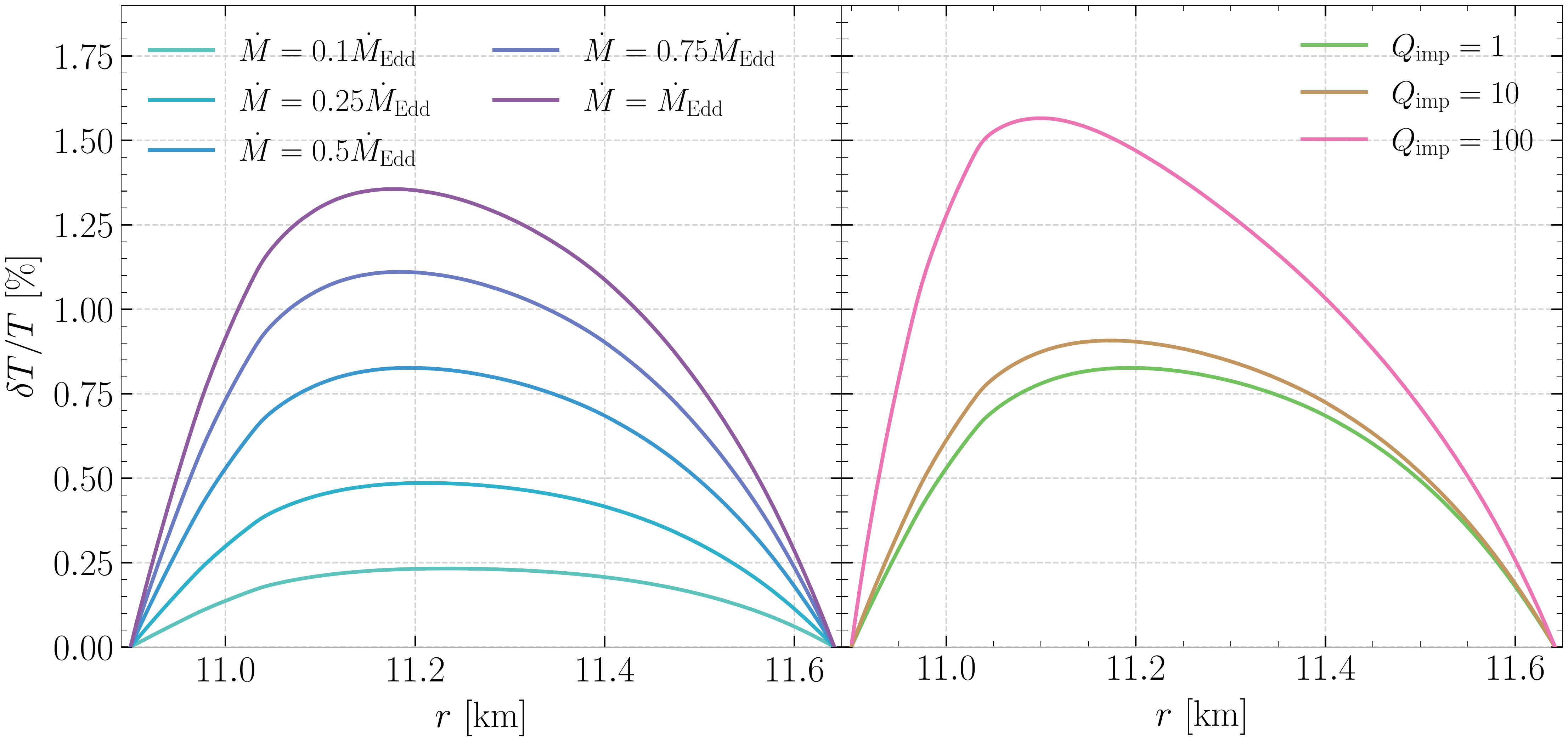}
    \caption{ Temperature perturbations in the crust for different accretion
    rates with a fixed impurity parameter  $Q_{\rm imp} =1 $  ({\it left}) and
    for impurity parameter with a fixed accretion rate $\dot{M} = 0.5
    \dot{M}_{\rm Edd} $ ({\it right}). Both panels have $f_{\rm nuc}=0.1$,
    $f_{\rm comp}=0$, and $\ell =2$.}
    \label{fig: PT_normal_1}
\end{figure*}

\begin{figure*}
    \centering 
    \includegraphics[width=12cm]{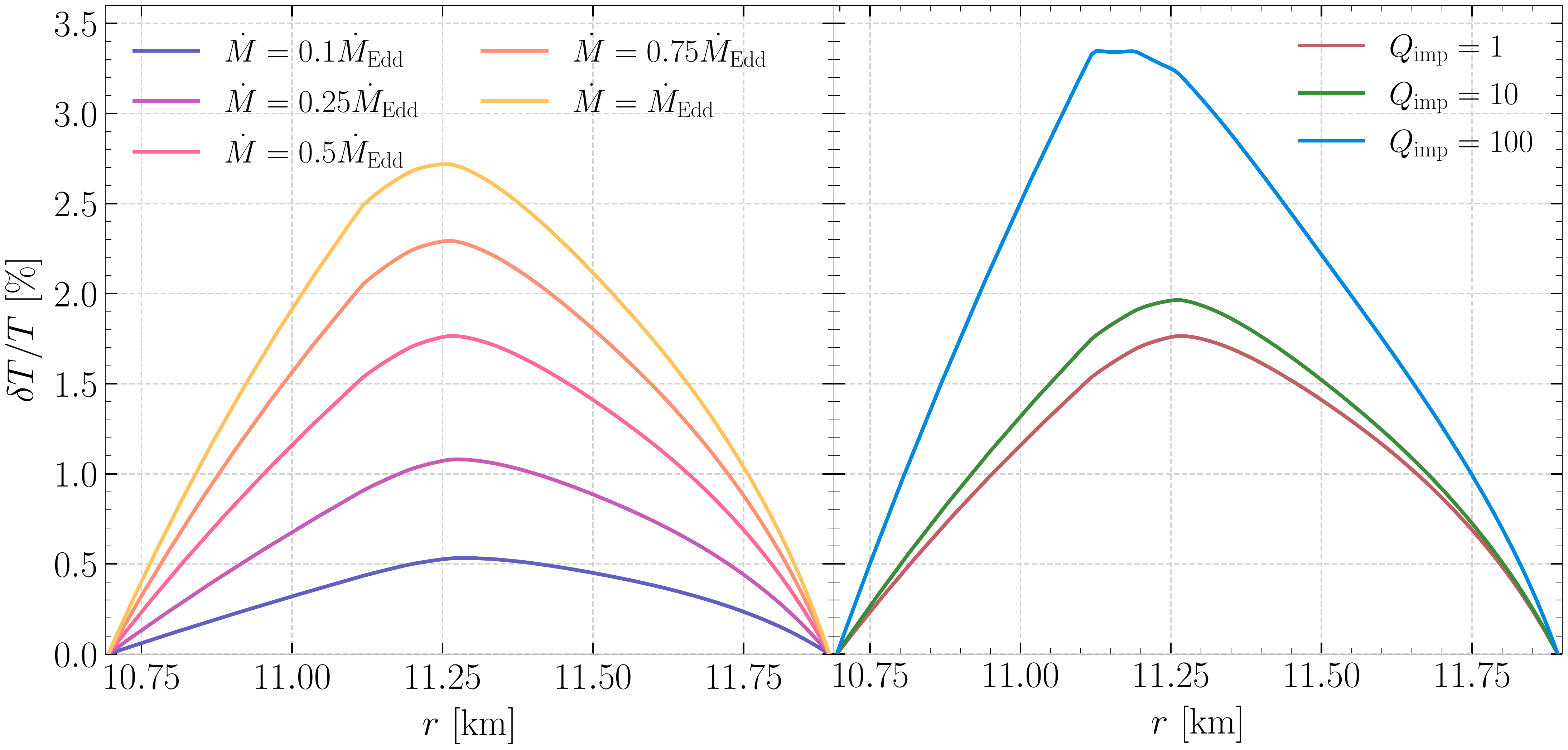}
    \caption{Same as Fig.~\ref{fig: PT_normal_1}, but for the effect of nuclear
    pasta.}
    \label{fig: PT_pasta_1}
\end{figure*} 

In the left panel of  Fig~\ref{fig: PT_normal_1}, we plot the temperature
perturbations as a function of the radius for different accretion rates. The
results from the lateral variations of nuclear heating, with $f_{\rm nuc}=0.1$
and  $f_{\rm comp}=0$, indicate that the conductivity and neutrino emission are
unperturbed. We have set $Q_{\rm imp} =1 $  and  $\ell =2$. The typical values
of $\delta T/T$ are a few times $1\%$ in the crust.  
In the right panel of  Fig.~\ref{fig: PT_normal_1}, we show the temperature
perturbations versus the radius for different impurity parameters. As the
impurity parameter increases, the temperature perturbations become larger.

To better understand the effect of nuclear pasta, we give our results in
Fig.~\ref{fig: PT_pasta_1}.  The size of the temperature perturbations increases
with higher accretion rates $\dot{M}$ and impurity parameters $Q_{\rm imp}$. The
maximal values of  $\delta T/T$ increase by a factor of two. 

\section{The elastic deformation of the crust}
\label{sec: elastic deformation}

When a star deforms from a perfect sphericity, it develops multipole moments.
These are defined as 
\begin{equation}\label{eq: Multipole}
    Q_{\ell m} \equiv \int_0^R \delta \rho_{\ell m}(r) r^{l + 2}{{\rm d} r} \,,
\end{equation}
where $(l, m)$ denotes the harmonic mode of the density perturbation $\delta
\rho_{\ell m}(r)$, and $R$ is the stellar radius. The quadrupole moment with
$(\ell, m) = (2, 2)$ contributes to the CGW signal. \citet{Bildsten:1998ey}
pointed out that the spin-up torque from accretions is balanced by the emission
of quadrupolar gravitational radiation. Under the condition for torque balance,
the CGWs amplitude can be written as \citep{Ushomirsky:2000ax}
\begin{equation}\label{eq: strain}
   h_{0} = \frac{32 \pi}{5} \left(\frac{\pi}{3}\right)^{1/2}\frac{G}{c^4}
   \frac{Q_{22}\, f^2_{\rm GW}}{d} \,,
\end{equation}
where $G$ and $c$ are the speed of light and the gravitational constant,
respectively, $d$ is the distance of the source, and the gravitational wave (GW)
frequency is $ f^2_{\rm GW} = 2 \nu$ for a NS rotating with spin frequency $\nu$
\citep{LIGOScientific:2019yhl, Reed:2021scb}.  


\subsection{The crustal perturbation equations}
\label{section: CPEqs}

Following \citet{Ushomirsky:2000ax}, the perturbed Euler equation for NSs with
an elastic crust in the Cowling approximation is written as
\begin{equation}\label{eq: elastic_per}
0 = - \nabla_i \delta P - \delta \rho \nabla_i \Phi + \nabla^j t_{i j} \,,
\end{equation}
where  $\Phi$ is the gravitational potential, and the shear-stress tensor $t_{i
j}$ describes elastic forces as a first-order perturbation. In our calculation,
the shear-stress tensor $t_{i j}$ can be written as 
\begin{equation}
    t_{i j} = \mu \left( \nabla_i \xi_j + \nabla_j \xi_i - \frac{2}{3} g_{i j}
    \nabla_k \xi^k \right) \,,
\end{equation}
where $g_{i j}$ is the flat three-metric, and the shear modulus $\mu$ in the
crust is,
\begin{equation}\label{eq: mu}
\mu^{\rm crust} = 0.1194 \frac{n_{\rm ion} (Ze)^2}{a}\,,
\end{equation}
where $a=[3/(4\pi n_{\rm ion})]^{1/3}$ is the average ion spacing. 
In particular, for the elastic properties of the nuclear pasta that was
described by \citet{Caplan:2018gkr}, \citet{Pethick:2020aey}, and
\citet{Xia:2022rhx}, the shear modulus could be $\mu^{\rm pasta} \approx
10^{30}\, \rm erg\, cm^{-3}$.

Employing tensor spherical harmonics leads to a convenient decomposition of the
shear-stress tensor \citep{Ushomirsky:2000ax, Haskell:2006sv},
\begin{align}\label{eq: shear-stress tensor}
t_{i j} =& t_{rr}Y_{\ell m}(\nabla_i r \nabla_j r -\frac{1}{2}e_{i j})+t_{r
\perp}(r)f_{i j} \nonumber \\
&+ t_{\Lambda}(r)(\Lambda_{i j}+\frac{1}{2}Y_{\ell m}e_{i j}) \,, 
\end{align}
where $t_{rr}$, $t_{r \perp}$, and $ t_{\Lambda}$ are functions of radial
coordinate $r$, $\beta = \sqrt{\ell (\ell + 1)}$ \citep{Ushomirsky:2000ax, Haskell:2006sv},  
\begin{eqnarray}
e_{i j} & = & g_{i j}- \nabla_i r \nabla_j r  \,, \\
f_{i j}  & = & \frac{r}{\beta} (\nabla_i r \nabla_j Y_{\ell m} + \nabla_j r
\nabla_i Y_{\ell m}) \,,\\
\Lambda_{i j } & = & \left(\frac{r}{\beta} \right)^2 \nabla_i  \nabla_j Y_{\ell
m}+ \frac{1}{\beta}f_{i j} \,.
\end{eqnarray}
The displacement vector $\xi^i$ is associated with  polar perturbations, and it
is given by,
\begin{equation}
    \xi^i = \xi_{r}(r) \nabla^{i} r Y_{\ell m} + \frac{r}{\beta} \xi_{\perp}(r)
    \nabla^{i} Y_{\ell m} \,,
\end{equation}
where $\xi_{r}(r)$ and $\xi_{\perp}(r)$ are the radial and tangential components
of the displacement, respectively.  Again, the perturbed continuity equation is
determined as 
\begin{equation}\label{eq: Continuity}
    \delta \rho =  - \rho  \frac{\partial \xi_{r}}{\partial r} - \left( \frac{2
    \rho}{r} + \frac{\partial \rho}{\partial r}  \right) \xi_{r} + \beta \rho
    \frac{\xi_{\perp}}{r}  \,. 
\end{equation} 

Using four new variables \citep{Ushomirsky:2000ax}
\begin{align}
    & z_{1} = \frac{\xi_{r}}{r} \label{eq: variables_1} \,, \\
    & z_{2} =\frac{t_{r r}}{P} - z_{1} {\frac{{\rm d} \ln P}{{\rm d} \ln r}}
    \label{eq: variables_2} \,, \\
    & z_{3} =\frac{\xi_{\perp}(r)}{\beta r}  \label{eq: variables_3}\,, \\
    & z_{4} = \frac{t_{r \perp}}{\beta P}  \label{eq: variables_4} \,, 
\end{align}
we obtain the following set of coupled ordinary differential equations (ODEs) \citep{Ushomirsky:2000ax},
\begin{align}
\frac{{\rm d} z_{1}}{{\rm d} \ln r}& = - \left( 1 + 2
\frac{\alpha_{2}}{\alpha_{3}} - \frac{\alpha_{4}}{\alpha_{3}}   \right)  z_{1} +
\frac{1}{\alpha_{3}}z_{2}  \nonumber\\ 
 & \quad  + \ell (\ell + 1)\frac{\alpha_{2}}{\alpha_{3}}z_{3} +
 \frac{1}{\alpha_{3}}\Delta S   \,, \label{eq: bc_p1} \\
\frac{{\rm d} z_{2}}{{\rm d} \ln r}&=\left( U V - 4 V + 12 \Gamma
\frac{\alpha_{1}}{\alpha_{3}} - 4 \frac{\alpha_{1} \alpha_{4}}{\alpha_{3}}
\right) z_{1}  \nonumber\\ 
 & \quad   + \left( V - 4  \frac{\alpha_{1}}{\alpha_{3}} \right) z_{2} + \ell
 (\ell +1 ) \left(V-6\Gamma  \frac{\alpha_{1}}{\alpha_{3}} \right) z_{3}
 \nonumber\\ 
  & \quad  +  \ell (\ell +1)z_{4} - 4 \frac{\alpha_{1}}{\alpha_{3}}\Delta S \,, 
  \label{eq: bc_p2} \\
\frac{{\rm d} z_{3}}{{\rm d} \ln r}&=- z_{1} + \frac{1}{\alpha_{1}}z_{4}  \,,
\label{eq: bc_p3}\\
\frac{{\rm d} z_{4}}{{\rm d} \ln r}&= \left( V - 6 \Gamma
\frac{\alpha_{1}}{\alpha_{3}} +2
\frac{\alpha_{1}\alpha_{4}}{\alpha_{3}}\right)z_{1} -
\frac{\alpha_{2}}{\alpha_{3}} z_{2} \nonumber\\ 
 & \quad  + \frac{2}{\alpha_{3}}\Big\{ [2\ell(\ell+1) -1] \alpha_{1}\alpha_{2} +
 2[\ell (\ell+1) -1]\alpha^2_{1} \Big\}z_{3}  \nonumber\\ 
  & \quad + (V -3 )z_{4} + 2  \frac{\alpha_{1}}{\alpha_{3}} \Delta S  \,,
  \label{eq: bc_p4} 
\end{align}
 where 
\begin{eqnarray}
U \equiv \frac{{\rm d} \ln g}{{\rm d} \ln r} +2 \,, \quad 
V \equiv \frac{\rho g r}{P} = - \frac{{\rm d} \ln P}{{\rm d} \ln r}  \,,  \quad
\quad \\ \nonumber
\alpha_{1}  \equiv  \frac{\mu}{P} \,, \quad 
\alpha_{2} \equiv  \Gamma - \frac{3}{2}\frac{\mu}{P} \,,  \quad
\alpha_{3}  \equiv \Gamma + \frac{4}{3} \frac{\mu}{P} \,,
\end{eqnarray}
where $g$ is the positive Newtonian gravitational acceleration.  For the  case
of $\delta T$ source term,  we have $\Gamma = \left. ({\partial \ln P}/{\partial
\ln \rho})\right|_{T}$, $\alpha_{4} = \left. ({\partial \ln P}/{\partial \ln T})
\right|_{\rho} ({{\rm d} \ln T}/{{\rm d} \ln r})$, and $\Delta S =
\left.({\partial \ln P}/{\partial \ln T})\right|_{\rho} ({\delta T}/{T})$.

We use the same boundary conditions as in  \citet{Ushomirsky:2000ax}.
At the interface between the core and crust, or nuclear pasta, we have 
\begin{equation}
z_{2} = Vz_{1} \,,  \quad z_{4} = 0 \,.
\end{equation}
However, at the interface between the crust and the ocean, the density may have
discontinuity, and we obtain the following boundary conditions, 
\begin{equation}
z_{2} = V \frac{\rho_{1}}{\rho_{\rm s}}z_{1} \,,  \quad z_{4} = 0 \,,
\end{equation}
where $(\rho_{\rm s} - \rho_{1})/\rho \approx 5 \times 10^{-5}$, $\rho_{\rm s} $
and $\rho_{1}$ are densities of elastic and liquid at the interface.
Furthermore, \citet{Gittins:2020cvx} constructed a new scheme for NS mountains
and modified the boundary conditions for fluid-elastic interfaces. 

When we discuss the density discontinuity at the fluid-elastic interfaces, the perturbed multipole moments can be written as \citep{Ushomirsky:2000ax}
\begin{eqnarray}\label{eq: Q_2} 
Q_{lm} =&-& \int^{r_{\rm top}}_{r_{\rm bot}}
  \frac{1}{V} \bigg \{ \ell(\ell+1)z_{4} - 2 \alpha_{1}\left( 2\frac{{\rm d}
  z_{1}}{{\rm d} \ln r} + \ell(\ell+1)z_{3} \right) \nonumber \\ 
	&+& \left(\ell+4 - U \right) \left(z_{2} - V z_{1} \right) \bigg \}
	r^{\ell+2}\rho \, {\rm d} r \,,
\end{eqnarray}
where $r_{\rm top}$ and $r_{\rm bot}$ are the top and bottom of the crust. For the case of nuclear pasta,  $r_{\rm bot}$ is the base of the nuclear pasta region. In the present work, we calculate the quadrupole moment, $Q_{22}$, using the above equation rigorously.

According to the results of  \citet{Ushomirsky:2000ax},  the fiducial quadrupole
moment $Q_{\text{fid}}$ generated by an individual capture layer  in the outer
crust can be approximated  as
\begin{multline}\label{eq: fiducial_outer}
  Q^{\rm outer}_{\rm{fid}} \approx 1.3 \times 10^{36} \, \rm{g \,cm}^{2} \, \\
   \times  R^4_6 \, \left( \frac{T}{10^8 \, \text{K}} \right) \, \left(
   \frac{\delta T /T }{1\%} \right) \, \left( \frac{E_{\text{cap}}}{30 \,
   \text{MeV}} \right)^3 \,,
\end{multline}
where $R_6 = R/10^{6}\, \rm cm$, and $E_{\text{cap}}$ is the threshold energy,
equivalent to the electron chemical potential $\mu_{e}$ at the transition
between one nuclear species and the next.  In the inner crust, the fiducial
quadrupole moment is defined as 
\begin{equation}\label{eq: fiducial_inner}
  Q^{\rm inner}_{\rm{fid}}  \equiv \int \left.\frac{\partial \ln \rho}{\partial
  \ln T}\right|_{P} \frac{\delta T}{T} \rho r^{\ell+2} {\rm d}r \,.
\end{equation}
In particular, we can calculate the fiducial value of the quadrupole moment for nuclear pasta using Eq.~(\ref{eq: fiducial_inner}). The fiducial value is $6.89 \times 10^{39}\, \rm g \,
cm^2$. This value is equivalent to $1.32 \times 10^{-6}\, \rm MR^2$, where $M =
1.84 \, M_{\odot}$ and $R = 11.89\, \rm km$.

\subsection{Results: elastic deformation}
\label{sec: ED_Results}

\begin{figure}
    \centering 
    \includegraphics[width=8cm]{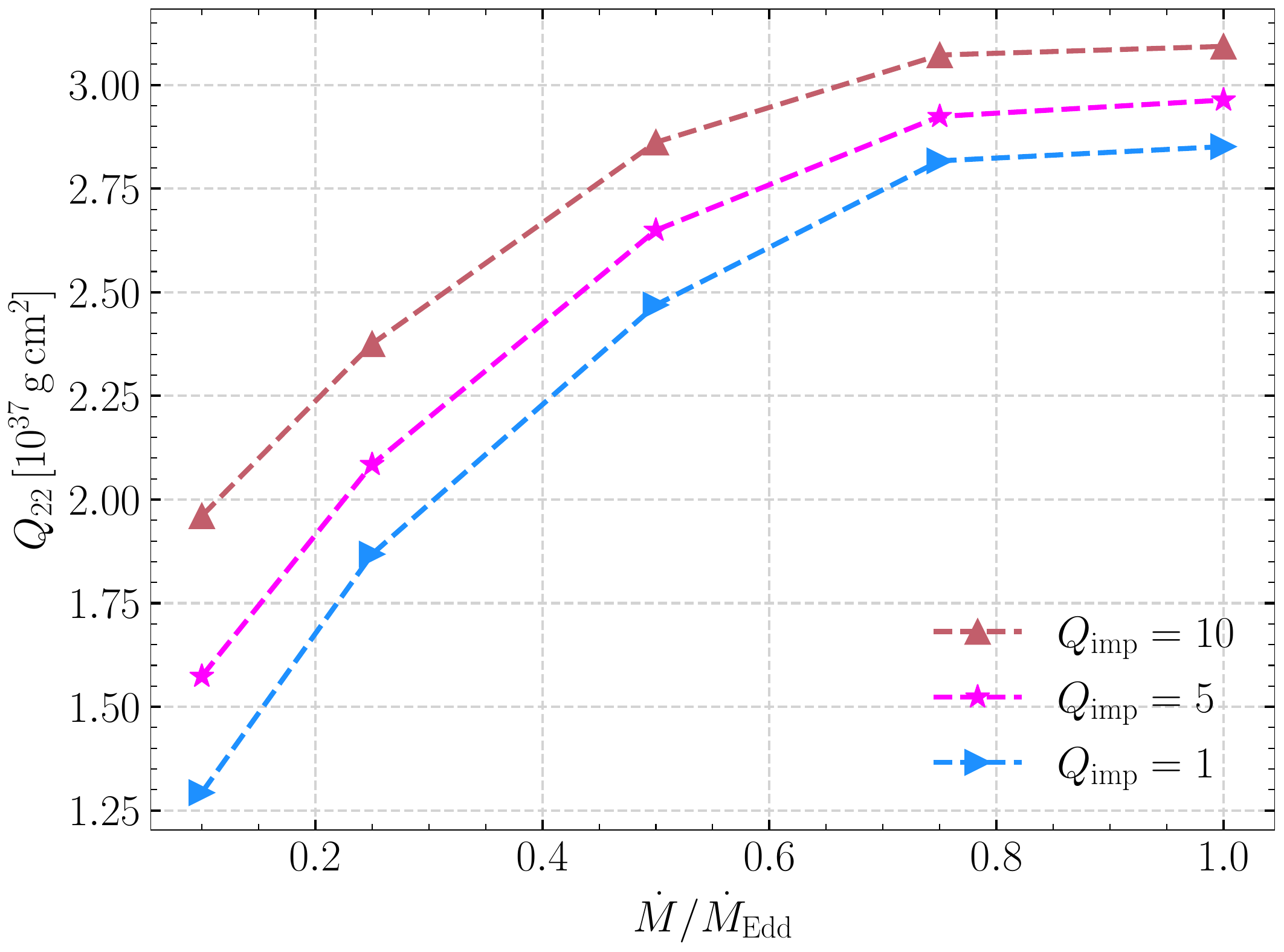}
    \caption{The absolute value of quadrupole moment as a function of the
    accretion rates without nuclear pasta for different impurity parameters.  We
    set  $f_{\rm nuc}=0.1$ and  $f_{\rm comp}=0$. }
    \label{fig: Q_22_crust}
\end{figure}

In Fig.~\ref{fig: Q_22_crust}, we show the absolute value of quadrupole moment
as a function of the accretion rates without nuclear pasta for different
impurity parameters. The quadrupole moment increases with the increase of the
accretion rates. The increase in impurity parameters leads to a higher
quadrupole moment. We interpret this as follows. The quadrupole moment is
directly proportional to the temperature perturbation, which is associated with
an increase in impurity parameter (see Eq.~(\ref{eq: fiducial_inner}) and right
panel of Fig.~\ref{fig: PT_pasta_1}).  According to our results, one finds that
the quadrupole moment without nuclear pasta is 
\begin{equation}\label{eq: results_1}
  Q^{\rm crust}_{22} \approx 2.2 \times 10^{37} \, \rm g\, cm^2 \,.
\end{equation}
The results are in order-of-magnitude agreement with the analytical results of
\citet{2024arXiv240700162J}. 

\begin{figure}
    \centering 
    \includegraphics[width=8cm]{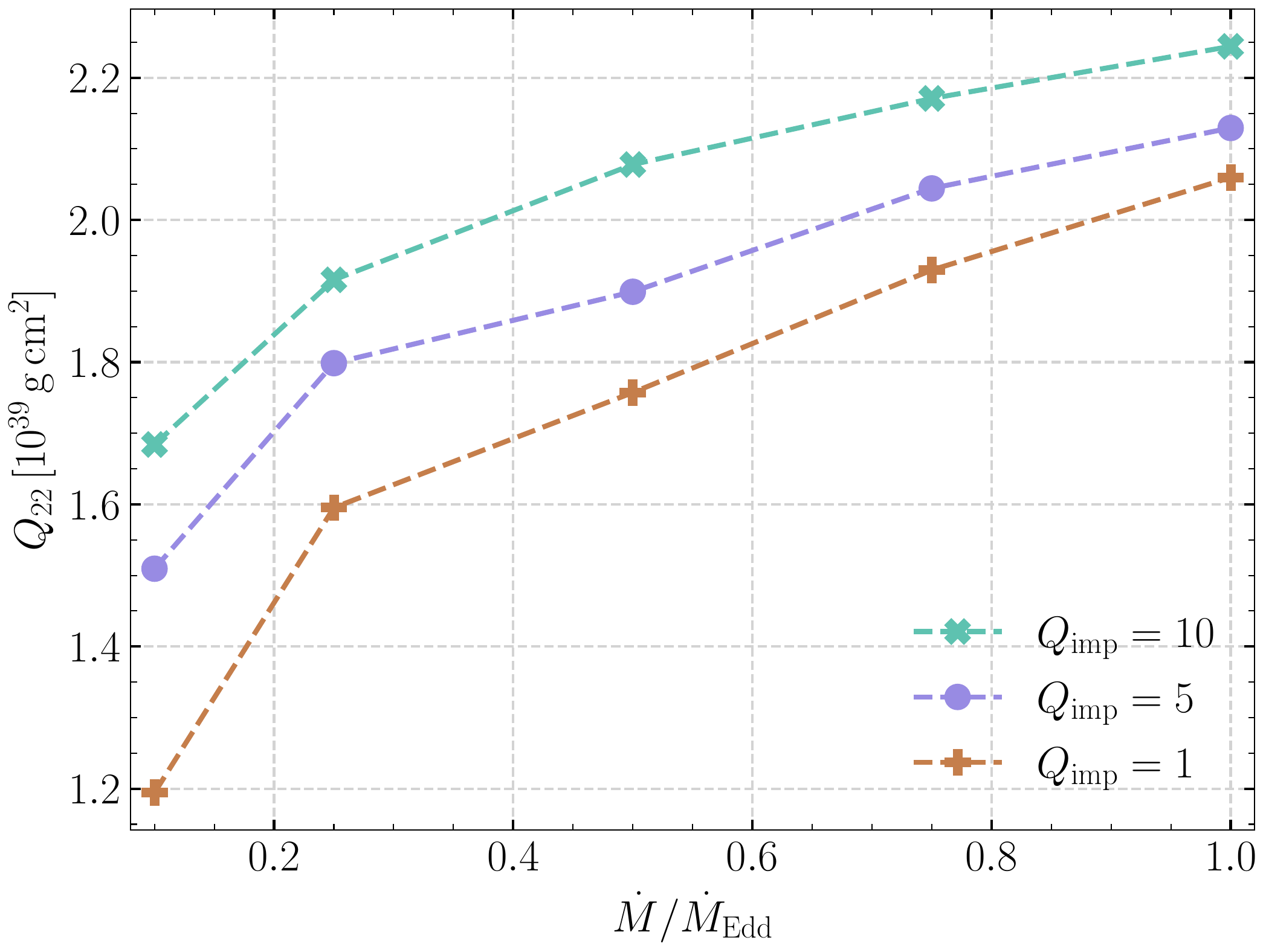}
    \caption{Same as Fig.~\ref{fig: Q_22_crust}, but for the effect of nuclear
    pasta.}
    \label{fig: Q_22_pasta}
\end{figure}

For the nuclear pasta model, we show in Fig.~\ref{fig: Q_22_pasta} the relation
of the quadrupole moment against accretion rates. One notices that this value of
the quadrupole moment is up to two orders of magnitude larger than the above
case. The quadrupole moment with nuclear pasta can be approximated by
\begin{equation}\label{eq: results_2}
  Q^{\rm pasta}_{22} \approx 1.7  \times 10^{39} \, \rm g\, cm^2 \,.
\end{equation}

\begin{table*}
    \centering
    \caption{Estimated spin frequency $\nu$, distance, outburst duration, and
    the long-term flux of different LMXBs. Sources in the top half of the table
    are AMXPs, while those in the bottom half are NXPs.}
    \renewcommand\arraystretch{1.25}
    \begin{tabular}{c c c c c c c c c c}
    \hline
    \hline
    Source &  $\nu$ & Distance & Outburst duration  & Long-term flux $F_{\rm av}$     & Reference  \\
                & (Hz)         &  (kpc)      &  (days)   & ($\times 10^{-8} \, \rm erg\,cm^{-2}\, s^{-1}$)  &          \\
    \hline
    AMXPs \\
    SAX J1808.4$-$3658      & 401      & 3.5       & 30  & $8.6 \times 10^{-3}$ & \citet{Patruno:2008km} \\
    XTE J1751$-$305           & 435      & 7.5      & 10  & $2.0 \times 10^{-3}$ & \citet{Miller:2002fh} \\
    XTE J1814$-$338           & 314      &  8        & 60  & $ 1.3 \times 10^{-3}$ & \citet{Haskell:2015psa} \\
    HETE J1900.1$-$2455   & 377     & 5           & 3000   & $1.8 \times 10^{-2}$  & \citet{Papitto:2013hza}  \\
    Aql X$-$1            & 550      & 5          & 30  & $1.2 \times 10^{-1}$ & \citet{Gungor:2014uia}  \\
    IGR J00291$+$5934       & 599      & 5          & 14   & $1.8 \times 10^{-3}$ & \citet{Falanga:2005th} \\
    \hline
    NXPs \\
    4U 1608$-$52                & 620     & 3.6       & 700    & $2.5 \times 10^{-1}$ &\citet{Gierlinski:2002ep}  \\
    4U 1636$-$536              & 581     & 5          & --       & $4.7 \times 10^{-1}$      &\citet{Haskell:2015psa}  \\
    4U 1702$-$429              & 329     & 5.5       & --       & $1.3 \times 10^{-1}$     &\citet{Haskell:2015psa}  \\
    4U 0614$+$091              & 415     & 3.2       & --       &$1.2 \times 10^{-1}$  &\citet{Piraino:1999qp} \\
    4U 1728$-$34               & 363      &5           & --       & $2.3 \times 10^{-1}$      &\citet{Egron:2011hy} \\
    KS 1731$-$260             & 526      &7           &4563      & $2.2 \times 10^{-1}$&\citet{Narita:2000zn} \\
    \hline 
    \end{tabular}
    \label{tab: Table_1}
\end{table*}

Next, we discuss the detectability of known accreting NSs using the data
available on the properties of the accreting NSs in LMXBs. We mainly consider
two relevant categories of accreting NSs: AMXPs and the nuclear-powered X-ray
pulsars  \citep[NXPs;][]{Watts:2008qw, Haskell:2015psa}. The details of the
LMXBs we use are listed in Table~\ref{tab: Table_1}. 

In addition, a fully coherent search over time is sensitive to a strain
\citep{Jaranowski:1998qm}
\begin{equation}\label{eq: Strain_2}
  h_{0} \approx 11.4 \sqrt{S_{n} (\nu) / T_{\rm obs}} \,,
\end{equation}
where $S_{n}(\nu) $ is the detector noise power spectral density, and the factor
$11.4$ accounts for a signal-to-noise ratio threshold, resulting in a single
trial false alarm rate of  $1$\% and a false dismissal rate of $10$\%
\citep{LIGOScientific:2006jsu, Watts:2008qw, Haskell:2015psa, 2024MNRAS.532.4550Z}. Detector network
sensitivities are shown for advanced LIGO (adv LIGO),  advanced Virgo \citep[adv
Virgo;][]{KAGRA:2013rdx},  Cosmic Explorer
\citep[CE;][]{LIGOScientific:2016wof}, and  Einstein Telescope
\citep[ET;][]{Punturo:2010zz, Maggiore:2019uih} \footnote{Strain sensitivity
data were taken from the publicly available LIGO document at
\url{https://dcc.ligo.org/LIGO-T1500293/public}.}.

\begin{figure}
    \centering 
    \includegraphics[width=8cm]{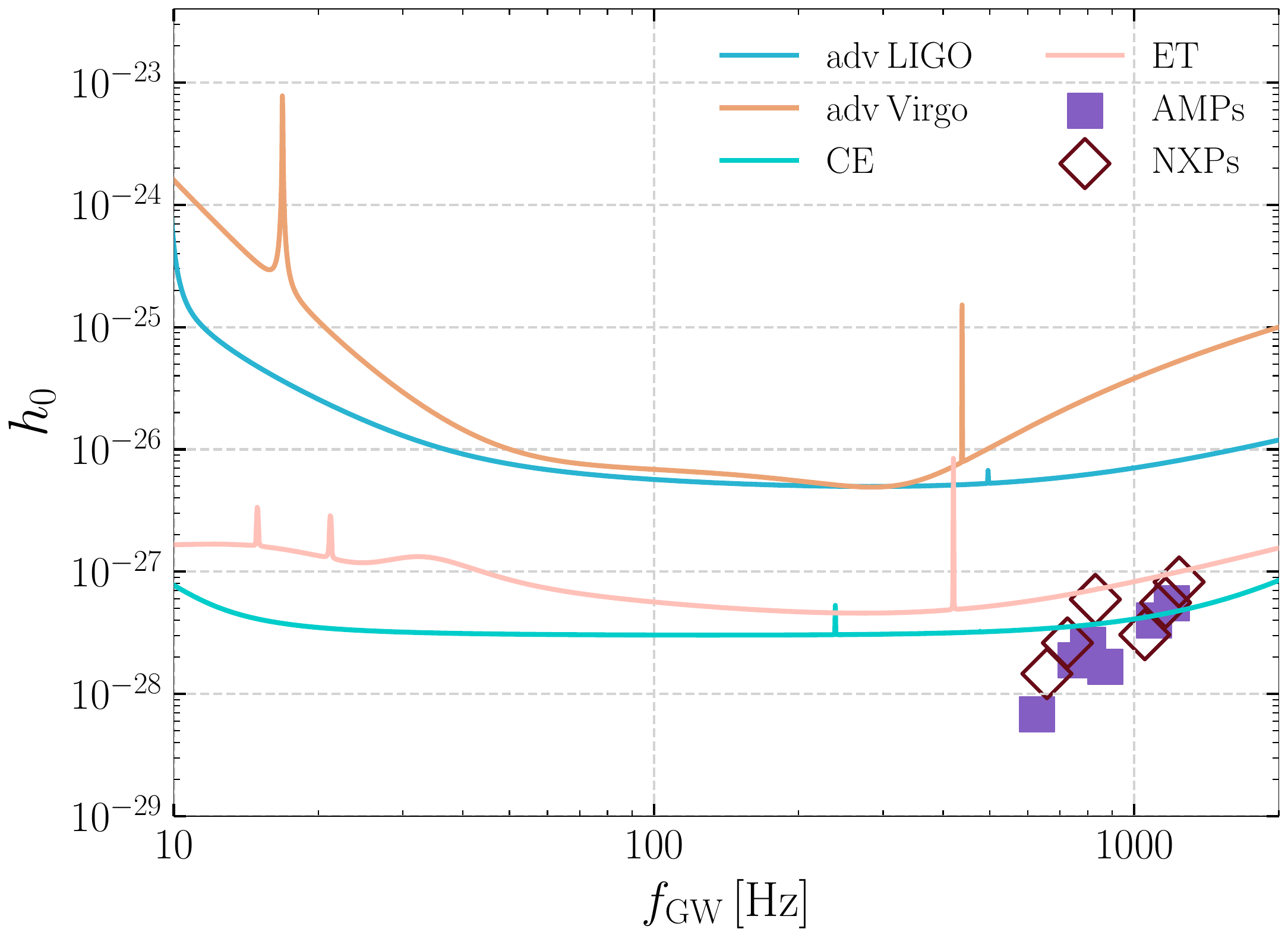}
    \caption{GW strain versus frequency for mountains in known pulsars. Closed
    squares denote AMXPs, and open diamonds indicate NXPs. The solid curves show
    the nominal strain noise sensitivity for various GW detectors.  The
    predicted results are from Eqs.~(\ref{eq: strain}) and~(\ref{eq:
    results_1}). }
    \label{fig: Strain_1}
\end{figure}

In Fig.~\ref{fig: Strain_1}, we compare predicted results to the detectable
amplitude limits. The detectable limits were obtained from Eq.~(\ref{eq:
Strain_2}) with $T_{\rm obs} = 2$ year, and our results are obtained from
Eq.~(\ref{eq: strain}) , where $Q_{22}$ is obtain from Eqs.~(\ref{eq:
results_1}) and~(\ref{eq: results_2}), respectively.  It is clear from the
figure that all systems fall well below the sensitivity curves for adv LIGO and
adv Virgo, with only a few sources promising detections by CE and ET.

\begin{figure}
    \centering 
    \includegraphics[width=8cm]{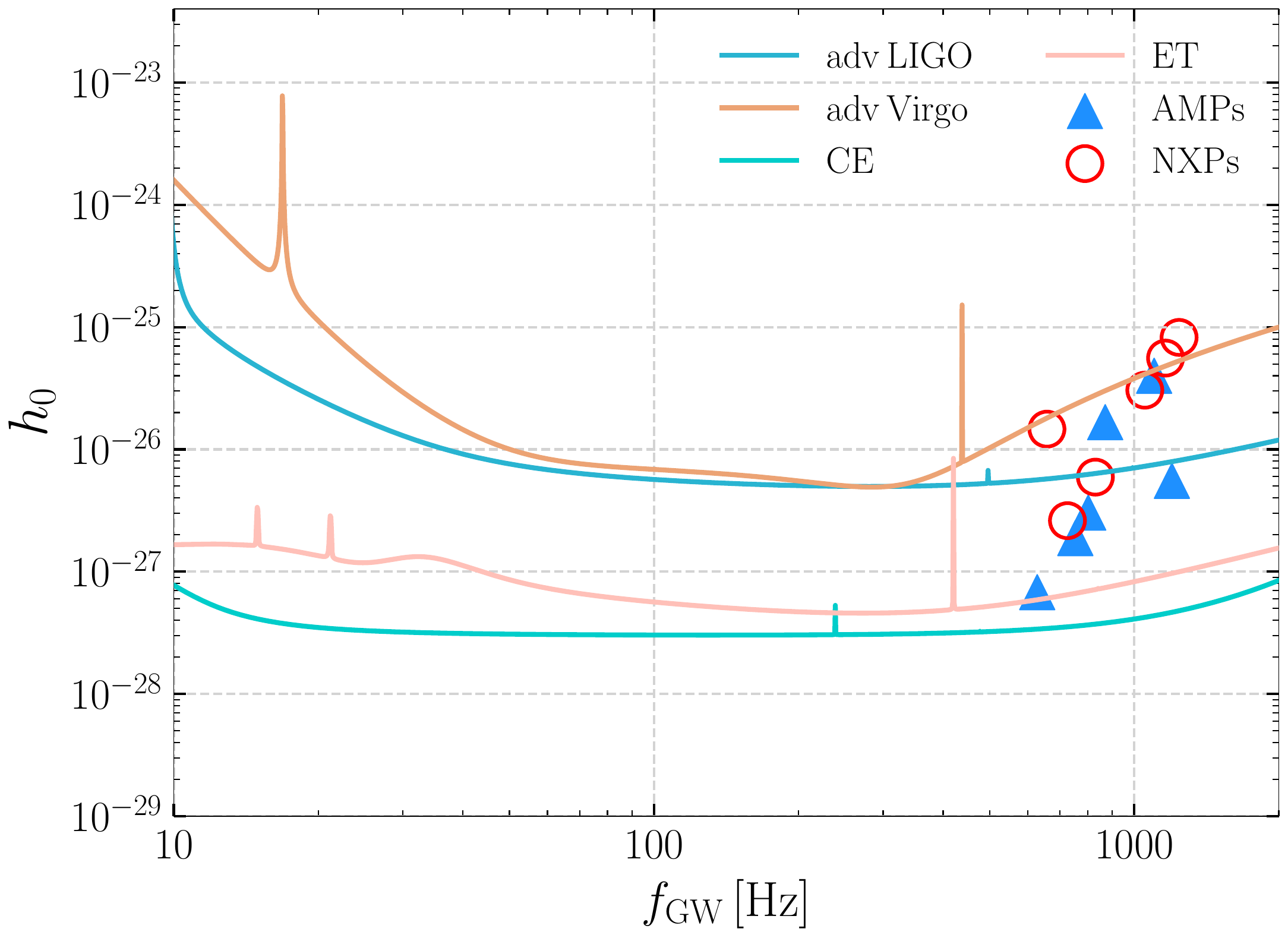}
    \caption{Similar to Fig.~\ref{fig: Strain_1}, but with the nuclear pasta
    effect. Closed triangles indicate AMXPs, and open circles indicate NXPs. The
    predicted results are from Eqs.~(\ref{eq: strain}) and~(\ref{eq:
    results_2}). }
    \label{fig: Strain_2}
\end{figure}

We next discuss the effect of nuclear pasta. Our results and the detectable
amplitudes are shown in Fig.~\ref{fig: Strain_2}. We can see that these sources
are well above the sensitivity curves for CE and ET detectors, indicating that
these sources are promising targets after considering the nuclear pasta phase.

\section{Breaking strain of the neutron star crust}
\label{sec: Breaking strain}

The breaking properties of an NS crust have implications for the CGWs
\citep{Horowitz:2009ya}, magnetar outbursts \citep{Thompson:2001ie}, resonant
crust shattering in NS mergers \citep{Tsang:2011ad}, and pulsar glitches
\citep{1969Natur.223..597R}.  Molecular dynamic simulations indicate the maximum
breaking stress around $0.1$  \citep{Horowitz:2009ya, Hoffman:2012nr}.  Other
estimates of the breaking strain approach to $0.04$ \citep{Baiko:2018jax}.
Following the approach in  \citet{Ushomirsky:2000ax}, we use the von Mises
criterion to define the elastic yield limit. The von Mises strain is
characterized by 
\begin{equation}\label{eq: von Mises strain}
\bar{\sigma}^2 \equiv \frac{1}{2}\sigma_{i j }\sigma^{i j } \,.
\end{equation}
Here  $\sigma_{i j } = t_{i j}/\mu$ is the strain tensor. We assume that the
crust will yield when $\bar{\sigma} \geq \sigma_{\rm max}$, where $\sigma_{\rm
max}$ is the breaking strain of the crust.  Using Eq.~(\ref{eq: shear-stress
tensor}), the right term of the above equation can be written as \citep{Ushomirsky:2000ax}
\begin{multline}\label{eq: shear-stress tensor_2}
\sigma_{i j }\sigma^{i j } = \frac{3}{2}\sigma^2_{r r}{\rm Re} \left(Y_{\ell m}\right)^2 
+ \sigma^2_{r \perp}{\rm Re} \left(f_{i j}\right)^2 \\ 
+ \sigma^2_{\Lambda} {\rm Re} \left( \Lambda_{i j} + \frac{1}{2}e_{i j}Y_{\ell m} \right)^2 \,.
\end{multline}
Here, recall that we define all variables to be real. We will calculate the
individual components of the shear strain using the expressions from
Eqs.~(\ref{eq: variables_1}) to  (\ref{eq: variables_4}). The individual
components of the breaking strain can be written as~\citep[see][for a detailed
variational derivation]{Ushomirsky:2000ax},
\begin{align}
   \sigma_{r r}(r)  &=   \frac{4}{3}\frac{{\rm d} z_{1}}{{\rm d} \ln r} +
   \frac{2}{3}\beta^2 z_{3} \,, \\
    \sigma_{r \perp}(r) & = \frac{\beta^2}{\mu} z_{4} \,, \\
   \sigma_{\Lambda}(r)  &=   2 \beta^2 z_{3} \,. 
\end{align}
Note that the $r r$-component of the shear strain at any point in the crust is
$\sigma_{r r}(r\,, \theta\,, \phi)  = \sigma_{r r}(r)\,{\rm Re}\{Y_{\ell
m}(\theta\,, \phi)\}$, similarly for the other components.  For illustration,
for $\ell = m =2$, the maximum value of angular factor is $(15/32\pi)^{1/2} =
0.39$ for $\sigma_{r r}$, $(5/8\pi)^{1/2} = 0.45$ for $\sigma_{r \perp}$, and
$(5/48\pi)^{1/2} = 0.18$ for $\sigma_{\Lambda}$ \citep{Ushomirsky:2000ax}.

\begin{figure}
    \centering 
    \includegraphics[width=8cm]{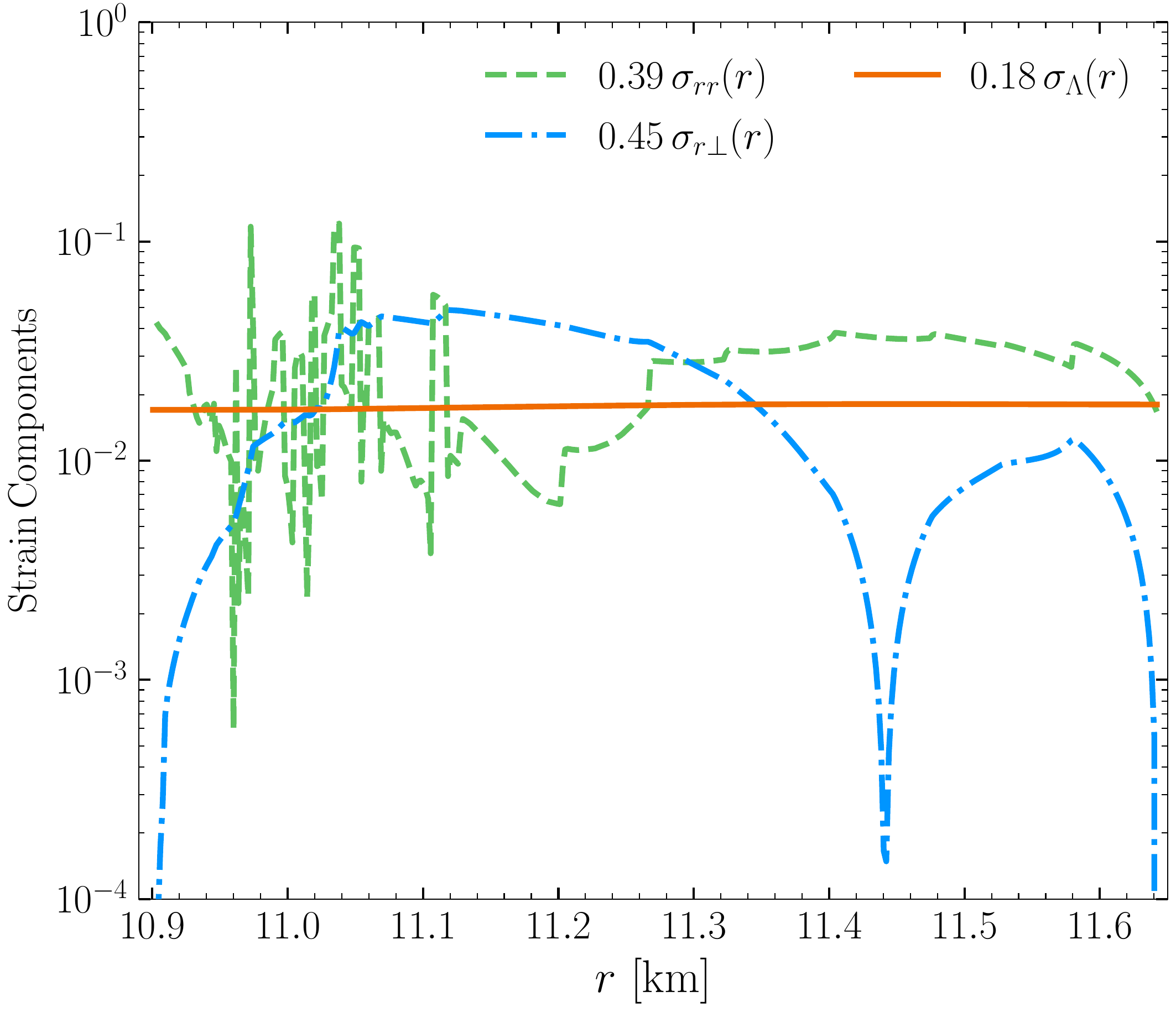}
    \caption{The maximum values of the different components of the shear strain
    without nuclear pasta as a function of the radius. The dashed line is
    $0.39\, \sigma_{r r}(r)$, the dashdot line denotes  $0.45\, \sigma_{r
    \perp}(r)$, and the solid line shows $0.18\,\sigma_{\Lambda}(r)$.  The
    background model has $\dot{M} = 0.5 \dot{M}_{\rm Edd} $, $f_{\rm nuc}=0.1$,
    and  $f_{\rm comp}=0$. }
    \label{fig: components_crust}
\end{figure}

\begin{figure}
    \centering 
    \includegraphics[width=8cm]{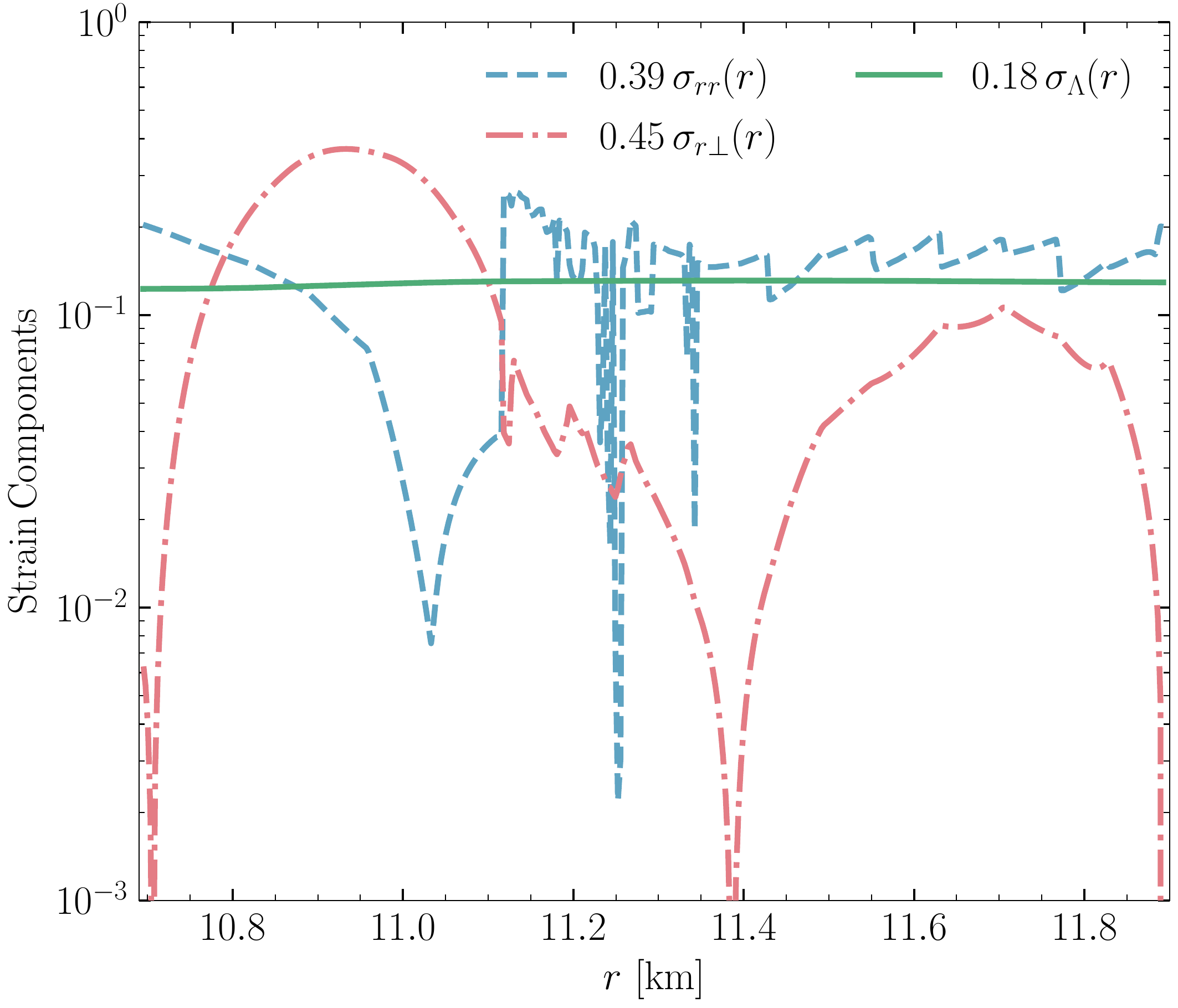}
    \caption{Same as Fig.~\ref{fig: components_crust}, but with the effect of
    the nuclear pasta.}
    \label{fig: components_pasta}
\end{figure}

In Fig. \ref{fig: components_crust}, we show the different components of the
shear strain without nuclear pasta.  A background model has  $\dot{M} = 0.5
\dot{M}_{\rm Edd} $, $f_{\rm nuc}=0.1$, and  $f_{\rm comp}=0$, and the results
of temperature perturbation are $\delta T /T= 0.85$\% with the maximum value of
the threshold energy $E_{\rm cap} = 43.69\, \rm MeV$. Note that the maximum
value of the threshold energy is dependent on the different EOS models
\citep{1990A&A...229..117H, Fantina:2018yad}. We can see that the $r
r$-component is larger than the other two components.  Moreover, the maximum
values of  $\sigma_{r \perp}(r)  $ and  $ \sigma_{\Lambda}(r) $  remain below
$0.1$. The results are similar to the results of \citet{Ushomirsky:2000ax}; see
Figs. 19 and 20 of their work. When $\sigma_{r r}(r)$ is the dominant stress
component, the expression of quadrupole moment becomes $Q^{\rm crust}_{22} = 2.2
\times 10^{37} \left(\sigma_{r r}/ 0.01 \right) \, \rm g\, cm^2$. 

Figure \ref{fig: components_pasta} presents the results illustrating the nuclear
pasta effect. We find that the maximum values of $\sigma_{r r}(r)$, $ \sigma_{r
\perp}(r) $, and $\sigma_{\Lambda}(r)$ range from $0.1$ to $0.4$, larger than
those in the previous case without nuclear pasta. Similarly, the expression for
the quadrupole moment becomes  $Q^{\rm pasta}_{22} = 1.7 \times 10^{39}
\left(\sigma_{i j}/ 0.1 \right) \, \rm g\, cm^2$, where $\sigma_{i j}$ can take
any of the three components. In particular, our results with nuclear pasta agree
with previous calculations of molecular dynamic simulations
\citep{Horowitz:2009ya, Hoffman:2012nr, Caplan:2018gkr}.  The large breaking
strain should support mountains on rapidly rotating NSs which are enough to
efficiently radiate CGW.
 
Note that the decrease in crust depth results in negative $\sigma_{r \perp}(r)
$. To compare with the values of  $\sigma_{r r}$ and $ \sigma_{\Lambda}(r) $, we
plot the absolute value of $\sigma_{r \perp}(r) $ in Figs. \ref{fig:
components_crust} and \ref{fig: components_pasta}.  

\section{Summary and discussion}
\label{sec: summary}

In this paper, we investigated the effect of nuclear pasta on the quadrupole
moments in LMXBs. We assume that the GW spin-down torque is balanced by
accretion-induced spin-up torque. The quadrupole moment needed to reach this
spin equilibrium is $Q_{22} = 1.7 \times 10^{39} \, \rm g\, cm^2$. Compared with
the results of \citet{Ushomirsky:2000ax}, the quadrupole moment is up to two
orders of magnitude [cf. Eq.~(\ref{eq: results_1})]. The main results of our
work are as follows. 

\begin{enumerate}[label=(\roman*)]

\item  By expanding the computational domain in our calculations, we have
introduced a more realistic representation of the thermal background of the
star. We have outlined the key mechanisms of neutrino emission and thermal
transport in the regions of the crust and nuclear pasta. Moreover, we discussed
the effects of impurity parameters on the temperature profiles and heat flux. 

\item Using the thermal perturbation equations, we have found that the
temperature variations can easily induce a required quadrupole moment with about
$10$\% ($f_{\rm nuc} = 0.1$) lateral variations of nuclear heating. We also
calculated the temperature variations versus different accretion rates and
impurity parameters in the nuclear pasta regions. In particular, the maximal
values of  $\delta T/T$ increase by a factor of two. 

\item Based on the crustal perturbation equations, we calculated the
relationship between the quadrupole moment and the accretion rates for both
regions of the crust and nuclear pasta. The quadrupole moments increase as the
accretion rates increase, which is the same as the case of the nuclear pasta
(see Figs.~\ref{fig: Q_22_crust} and \ref{fig: Q_22_pasta}). In particular, the
quadrupole moment is up to two orders of magnitude larger than the case without
nuclear pasta.

\item We studied the detectability of known accreting NSs, including AMXPs and
NXPs, and compared predicted results to the detectable amplitude limits. In the
case without nuclear pasta, we found all sources that we considered fall
significantly below the sensitivity curves for adv LIGO and adv Virgo detectors,
with only a few sources promising detections by CE and ET (see Fig.~\ref{fig:
Strain_1}). However, these sources are well above the sensitivity curves for CE
and ET detectors after considering the nuclear pasta phase. Because of the
nuclear pasta properties, these sources may be promising sources for
next-generation GW detectors (see Fig.~\ref{fig: Strain_2}).

\item We calculated the maximum shear strain for a NS crust.  The maximum value
for the quadrupole moment scales with the maximum strain as $Q^{\rm crust}_{\rm
max} = 2.2 \times 10^{37} \left(\sigma_{\rm max}/ 0.01 \right) \, \rm g\, cm^2$
(see Fig.~\ref{fig: components_crust}).  We found that, in the case of nuclear
pasta, the maximum values of the shear strain range from $0.1$ to $0.4$, which
are larger than those in the previous case. The general relation between the
maximum quadrupole moment and breaking strain is expressed as $Q^{\rm
pasta}_{\rm max} = 1.7 \times 10^{39} \left(\sigma_{\rm max}/ 0.1 \right) \, \rm
g\, cm^2$ (see Fig.~\ref{fig: components_pasta}).

\end{enumerate}

We next outline several areas where our model could be improved in future work. 
First, we used the EOS of  \citet{1990A&A...229..117H, 1990A&A...227..431H} to
calculated the thermal model. This could be extended to higher-density regions,
which are associated with the bottom boundary.  Second, we should adopt the
modern EOS for accreted crust, e.g. the one provided by \citet{Fantina:2018yad}.
Note that the total heat released per nucleon is probably not significantly
different from the $1.33 \, \rm MeV$ used in this work. For illustration, Using
a value of $1.5 \, \rm MeV$ per nucleon \citep{Fantina:2018yad,
Fantina:2022xas}, \citet{Hutchins:2022chj} investigated the effects of magnetic
fields on the thermal mountains.  Third, in calculating the temperature profiles
of the nuclear pasta, we have used the thermal conductivity independent of the
density (see Eq.~\ref{eq: conductivity_2}). However, as the density increases,
it leads to a larger thermal conductivity \citep{Horowitz:2008vf}, therefore, we
should consider the effect of the different thermal conductivities on the
temperature profiles and heat flux in future work.  Finally, in our analysis, we
employed the Cowling approximation, meaning we neglected the perturbation of the
gravitational potential.  According to the discussion by
\citet{Ushomirsky:2000ax}, the quadrupole moment will enhance the result by
$25$--$200$\% in the full self-consistent calculation.  One shall employ a more
rigorous treatment to compute the quadrupole moment in future work. 

In the present work, we have studied the effect of the nuclear pasta in
high-density regions on the background model, temperature perturbation,
quadrupole moment, and shear strain. However, \citet{2024arXiv240700162J}
pointed out that deep capture layers may not exist based on the EOS model of the
accreted crust of \citet{Gusakov:2020hij}. Hence, they discussed an alternative
source of temperature, coming from the crustal lattice pressure. Furthermore, an
anisotropic NS crust will produce non-axisymmetric deformations
\citep{Morales:2023euv} and support an ellipticity ranges from $10^{-9}$ to
$10^{-8}$ \citep{2024arXiv240914482M}. These can be considered in future work.

\section{Acknowledgements}\label{sec: Acknowledgements}

We thank Zexin Hu for helpful discussions.  This work was supported by the
National SKA Program of China (2020SKA0120300, 2020SKA0120100), the National
Natural Science Foundation of China (11991053, 12275234, 12342027), the Beijing
Natural Science Foundation (1242018), the Max Planck Partner Group Program
funded by the Max Planck Society, and the High-Performance Computing Platform of
Peking University. 

%

%


\bibliography{mountain}{}

\begin{thebibliography}{}
\expandafter\ifx\csname natexlab\endcsname\relax\def\natexlab#1{#1}\fi
\providecommand{\url}[1]{\href{#1}{#1}}
\providecommand{\dodoi}[1]{doi:~\href{http://doi.org/#1}{\nolinkurl{#1}}}
\providecommand{\doeprint}[1]{\href{http://ascl.net/#1}{\nolinkurl{http://ascl.net/#1}}}
\providecommand{\doarXiv}[1]{\href{https://arxiv.org/abs/#1}{\nolinkurl{https://arxiv.org/abs/#1}}}

\bibitem[{Abadie {et~al.}(2011)}]{LIGOScientific:2011msu}
Abadie, J., {et~al.} 2011, ApJ, 737, 93, \dodoi{10.1088/0004-637X/737/2/93}

\bibitem[{Abbott {et~al.}(2005)}]{LIGOScientific:2004sbr}
Abbott, B., {et~al.} 2005, Phys. Rev. Lett., 94, 181103, \dodoi{10.1103/PhysRevLett.94.181103}

\bibitem[{Abbott {et~al.}(2007{\natexlab{a}})}]{LIGOScientific:2007leh}
---. 2007{\natexlab{a}}, Phys. Rev. D, 76, 042001, \dodoi{10.1103/PhysRevD.76.042001}

\bibitem[{Abbott {et~al.}(2007{\natexlab{b}})}]{LIGOScientific:2006jsu}
---. 2007{\natexlab{b}}, Phys. Rev. D, 76, 082001, \dodoi{10.1103/PhysRevD.76.082001}

\bibitem[{Abbott {et~al.}(2016)}]{KAGRA:2013rdx}
Abbott, B.~P., {et~al.} 2016, Living Rev. Rel., 19, 1, \dodoi{10.1007/s41114-020-00026-9}

\bibitem[{Abbott {et~al.}(2017{\natexlab{a}})}]{LIGOScientific:2017csd}
---. 2017{\natexlab{a}}, Phys. Rev. D, 96, 062002, \dodoi{10.1103/PhysRevD.96.062002}

\bibitem[{Abbott {et~al.}(2017{\natexlab{b}})}]{LIGOScientific:2016wof}
---. 2017{\natexlab{b}}, Class. Quant. Grav., 34, 044001, \dodoi{10.1088/1361-6382/aa51f4}

\bibitem[{Abbott {et~al.}(2019)}]{LIGOScientific:2019yhl}
---. 2019, Phys. Rev. D, 100, 024004, \dodoi{10.1103/PhysRevD.100.024004}

\bibitem[{Abbott {et~al.}(2021)}]{KAGRA:2021una}
Abbott, R., {et~al.} 2021, Phys. Rev. D, 104, 082004, \dodoi{10.1103/PhysRevD.104.082004}

\bibitem[{Abbott {et~al.}(2022{\natexlab{a}})}]{LIGOScientific:2021ozr}
---. 2022{\natexlab{a}}, Phys. Rev. D, 105, 022002, \dodoi{10.1103/PhysRevD.105.022002}

\bibitem[{Abbott {et~al.}(2022{\natexlab{b}})}]{KAGRA:2022dwb}
---. 2022{\natexlab{b}}, Phys. Rev. D, 106, 102008, \dodoi{10.1103/PhysRevD.106.102008}

\bibitem[{Andersson {et~al.}(2005)Andersson, Glampedakis, Haskell, \& Watts}]{Andersson:2004zz}
Andersson, N., Glampedakis, K., Haskell, B., \& Watts, A.~L. 2005, MNRAS, 361, 1153, \dodoi{10.1111/j.1365-2966.2005.09167.x}

\bibitem[{Andersson {et~al.}(1999)Andersson, Kokkotas, \& Stergioulas}]{Andersson:1998qs}
Andersson, N., Kokkotas, K.~D., \& Stergioulas, N. 1999, ApJ, 516, 307, \dodoi{10.1086/307082}

\bibitem[{Baiko \& Chugunov(2018)}]{Baiko:2018jax}
Baiko, D.~A., \& Chugunov, A.~I. 2018, MNRAS, 480, 5511, \dodoi{10.1093/mnras/sty2259}

\bibitem[{Baiko {et~al.}(2001)Baiko, Haensel, \& Yakovlev}]{Baiko:2001cj}
Baiko, D.~A., Haensel, P., \& Yakovlev, D.~G. 2001, A\&A, 374, 151, \dodoi{10.1051/0004-6361:20010621}

\bibitem[{Bildsten(1998)}]{Bildsten:1998ey}
Bildsten, L. 1998, ApJL, 501, L89, \dodoi{10.1086/311440}

\bibitem[{Brown(2000)}]{Brown:1999dk}
Brown, E.~F. 2000, ApJ, 531, 988, \dodoi{10.1086/308487}

\bibitem[{Brown \& Cumming(2009)}]{Brown:2009kw}
Brown, E.~F., \& Cumming, A. 2009, ApJ, 698, 1020, \dodoi{10.1088/0004-637X/698/2/1020}

\bibitem[{Caplan {et~al.}(2021)Caplan, Forsman, \& Schneider}]{Caplan:2020ewl}
Caplan, M.~E., Forsman, C.~R., \& Schneider, A.~S. 2021, Phys. Rev. C, 103, 055810, \dodoi{10.1103/PhysRevC.103.055810}

\bibitem[{Caplan \& Horowitz(2017)}]{Caplan:2016uvu}
Caplan, M.~E., \& Horowitz, C.~J. 2017, Rev. Mod. Phys., 89, 041002, \dodoi{10.1103/RevModPhys.89.041002}

\bibitem[{Caplan {et~al.}(2018)Caplan, Schneider, \& Horowitz}]{Caplan:2018gkr}
Caplan, M.~E., Schneider, A.~S., \& Horowitz, C.~J. 2018, Phys. Rev. Lett., 121, 132701, \dodoi{10.1103/PhysRevLett.121.132701}

\bibitem[{Chakrabarty {et~al.}(2003)Chakrabarty, Morgan, Muno, Galloway, Wijnands, van~der Klis, \& Markwardt}]{Chakrabarty:2003kt}
Chakrabarty, D., Morgan, E.~H., Muno, M.~P., {et~al.} 2003, Nature, 424, 42, \dodoi{10.1038/nature01732}

\bibitem[{{Cook} {et~al.}(1994){Cook}, {Shapiro}, \& {Teukolsky}}]{1994ApJ...423L.117C}
{Cook}, G.~B., {Shapiro}, S.~L., \& {Teukolsky}, S.~A. 1994, ApJL, 423, L117, \dodoi{10.1086/187250}

\bibitem[{Deibel {et~al.}(2017)Deibel, Cumming, Brown, \& Reddy}]{Deibel:2016vbc}
Deibel, A., Cumming, A., Brown, E.~F., \& Reddy, S. 2017, ApJ, 839, 95, \dodoi{10.3847/1538-4357/aa6a19}

\bibitem[{Dorso {et~al.}(2020)Dorso, Strachan, \& Frank}]{Dorso:2020zhk}
Dorso, C.~O., Strachan, A., \& Frank, G.~A. 2020, Nucl. Phys. A, 1002, 122004, \dodoi{10.1016/j.nuclphysa.2020.122004}

\bibitem[{Douchin \& Haensel(2001)}]{Douchin:2001sv}
Douchin, F., \& Haensel, P. 2001, A\&A, 380, 151, \dodoi{10.1051/0004-6361:20011402}

\bibitem[{Egron {et~al.}(2011)}]{Egron:2011hy}
Egron, E., {et~al.} 2011, A\&A, 530, A99, \dodoi{10.1051/0004-6361/201016093}

\bibitem[{Falanga {et~al.}(2005)Falanga, Kuiper, Poutanen, Bonning, Hermsen, Di~Salvo, Goldoni, Goldwurm, Shaw, \& Stella}]{Falanga:2005th}
Falanga, M., Kuiper, L., Poutanen, J., {et~al.} 2005, A\&A, 444, 15, \dodoi{10.1051/0004-6361:20053472}

\bibitem[{Fantina {et~al.}(2018)Fantina, Zdunik, Chamel, Pearson, Haensel, \& Goriely}]{Fantina:2018yad}
Fantina, A.~F., Zdunik, J.~L., Chamel, N., {et~al.} 2018, A\&A, 620, A105, \dodoi{10.1051/0004-6361/201833605}

\bibitem[{Fantina {et~al.}(2022)Fantina, Zdunik, Chamel, Pearson, Suleiman, \& Goriely}]{Fantina:2022xas}
---. 2022, A\&A, 665, A74, \dodoi{10.1051/0004-6361/202243715}

\bibitem[{Fujisawa {et~al.}(2022)Fujisawa, Kisaka, \& Kojima}]{Fujisawa:2022dzp}
Fujisawa, K., Kisaka, S., \& Kojima, Y. 2022, MNRAS, 516, 5196, \dodoi{10.1093/mnras/stac2585}

\bibitem[{Galloway {et~al.}(2005)Galloway, Markwardt, Morgan, Chakrabarty, \& Strohmayer}]{Galloway:2005rg}
Galloway, D.~K., Markwardt, C.~B., Morgan, E.~H., Chakrabarty, D., \& Strohmayer, T.~E. 2005, ApJL, 622, L45, \dodoi{10.1086/429563}

\bibitem[{Gao {et~al.}(2023)Gao, Shao, Desvignes, Jones, Kramer, \& Yim}]{Gao:2022hzd}
Gao, Y., Shao, L., Desvignes, G., {et~al.} 2023, MNRAS, 519, 1080, \dodoi{10.1093/mnras/stac3546}

\bibitem[{Gao {et~al.}(2020)Gao, Shao, Xu, Sun, Liu, \& Xu}]{Gao:2020zcd}
Gao, Y., Shao, L., Xu, R., {et~al.} 2020, MNRAS, 498, 1826, \dodoi{10.1093/mnras/staa2476}

\bibitem[{{Ghosh} \& {Lamb}(1978)}]{1978ApJ...223L..83G}
{Ghosh}, P., \& {Lamb}, F.~K. 1978, ApJL, 223, L83, \dodoi{10.1086/182734}

\bibitem[{Gierlinski \& Done(2002)}]{Gierlinski:2002ep}
Gierlinski, M., \& Done, C. 2002, MNRAS, 337, 1373, \dodoi{10.1046/j.1365-8711.2002.06009.x}

\bibitem[{Gittins {et~al.}(2020)Gittins, Andersson, \& Jones}]{Gittins:2020cvx}
Gittins, F., Andersson, N., \& Jones, D.~I. 2020, MNRAS, 500, 5570, \dodoi{10.1093/mnras/staa3635}

\bibitem[{G\"ung\"or {et~al.}(2014)G\"ung\"or, G\"uver, \& Ek\c{s}i}]{Gungor:2014uia}
G\"ung\"or, C., G\"uver, T., \& Ek\c{s}i, K.~Y. 2014, MNRAS, 439, 2717, \dodoi{10.1093/mnras/stu128}

\bibitem[{Gusakov \& Chugunov(2020)}]{Gusakov:2020hij}
Gusakov, M.~E., \& Chugunov, A.~I. 2020, Phys. Rev. Lett., 124, 191101, \dodoi{10.1103/PhysRevLett.124.191101}

\bibitem[{Haensel {et~al.}(1996)Haensel, Kaminker, \& Yakovlev}]{Haensel:1996rd}
Haensel, P., Kaminker, A.~D., \& Yakovlev, D.~G. 1996, A\&A, 314, 328.
\newblock \doarXiv{astro-ph/9604073}

\bibitem[{Haensel {et~al.}(2007)Haensel, Potekhin, \& Yakovlev}]{Haensel:2007yy}
Haensel, P., Potekhin, A.~Y., \& Yakovlev, D.~G. 2007, {Neutron stars 1: Equation of state and structure}, Vol. 326 (New York, USA: Springer), \dodoi{10.1007/978-0-387-47301-7}

\bibitem[{{Haensel} \& {Zdunik}(1990{\natexlab{a}})}]{1990A&A...229..117H}
{Haensel}, P., \& {Zdunik}, J.~L. 1990{\natexlab{a}}, A\&A, 229, 117

\bibitem[{{Haensel} \& {Zdunik}(1990{\natexlab{b}})}]{1990A&A...227..431H}
---. 1990{\natexlab{b}}, A\&A, 227, 431

\bibitem[{Haskell {et~al.}(2006)Haskell, Jones, \& Andersson}]{Haskell:2006sv}
Haskell, B., Jones, D.~I., \& Andersson, N. 2006, MNRAS, 373, 1423, \dodoi{10.1111/j.1365-2966.2006.10998.x}

\bibitem[{Haskell {et~al.}(2015)Haskell, Priymak, Patruno, Oppenoorth, Melatos, \& Lasky}]{Haskell:2015psa}
Haskell, B., Priymak, M., Patruno, A., {et~al.} 2015, MNRAS, 450, 2393, \dodoi{10.1093/mnras/stv726}

\bibitem[{Haskell {et~al.}(2008)Haskell, Samuelsson, Glampedakis, \& Andersson}]{Haskell:2007bh}
Haskell, B., Samuelsson, L., Glampedakis, K., \& Andersson, N. 2008, MNRAS, 385, 531, \dodoi{10.1111/j.1365-2966.2008.12861.x}

\bibitem[{Heyl(2002)}]{Heyl:2002pe}
Heyl, J.~S. 2002, ApJL, 574, L57, \dodoi{10.1086/342263}

\bibitem[{Hoffman \& Heyl(2012)}]{Hoffman:2012nr}
Hoffman, K., \& Heyl, J. 2012, MNRAS, 426, 2404, \dodoi{10.1111/j.1365-2966.2012.21921.x}

\bibitem[{Horowitz \& Berry(2008)}]{Horowitz:2008vf}
Horowitz, C.~J., \& Berry, D.~K. 2008, Phys. Rev. C, 78, 035806, \dodoi{10.1103/PhysRevC.78.035806}

\bibitem[{Horowitz {et~al.}(2015)Horowitz, Berry, Briggs, Caplan, Cumming, \& Schneider}]{Horowitz:2014xca}
Horowitz, C.~J., Berry, D.~K., Briggs, C.~M., {et~al.} 2015, Phys. Rev. Lett., 114, 031102, \dodoi{10.1103/PhysRevLett.114.031102}

\bibitem[{Horowitz \& Kadau(2009)}]{Horowitz:2009ya}
Horowitz, C.~J., \& Kadau, K. 2009, Phys. Rev. Lett., 102, 191102, \dodoi{10.1103/PhysRevLett.102.191102}

\bibitem[{Hutchins \& Jones(2023)}]{Hutchins:2022chj}
Hutchins, T.~J., \& Jones, D.~I. 2023, MNRAS, 522, 226, \dodoi{10.1093/mnras/stad967}

\bibitem[{Jaranowski {et~al.}(1998)Jaranowski, Krolak, \& Schutz}]{Jaranowski:1998qm}
Jaranowski, P., Krolak, A., \& Schutz, B.~F. 1998, Phys. Rev. D, 58, 063001, \dodoi{10.1103/PhysRevD.58.063001}

\bibitem[{Jones \& Andersson(2002)}]{Jones:2001yg}
Jones, D.~I., \& Andersson, N. 2002, MNRAS, 331, 203, \dodoi{10.1046/j.1365-8711.2002.05180.x}

\bibitem[{{Jones} \& {Hutchins}(2024)}]{2024arXiv240700162J}
{Jones}, D.~I., \& {Hutchins}, T.~J. 2024, arXiv e-prints, arXiv:2407.00162, \dodoi{10.48550/arXiv.2407.00162}

\bibitem[{{Jones} \& {Riles}(2024)}]{2024arXiv240302066J}
{Jones}, D.~I., \& {Riles}, K. 2024, arXiv e-prints, arXiv:2403.02066, \dodoi{10.48550/arXiv.2403.02066}

\bibitem[{Lattimer \& Prakash(2007)}]{Lattimer:2006xb}
Lattimer, J.~M., \& Prakash, M. 2007, Phys. Rept., 442, 109, \dodoi{10.1016/j.physrep.2007.02.003}

\bibitem[{Levin(1999)}]{Levin:1998wa}
Levin, Y. 1999, ApJ, 517, 328, \dodoi{10.1086/307196}

\bibitem[{Lin {et~al.}(2020)Lin, Caplan, Horowitz, \& Lunardini}]{Lin:2020nxy}
Lin, Z., Caplan, M.~E., Horowitz, C.~J., \& Lunardini, C. 2020, Phys. Rev. C, 102, 045801, \dodoi{10.1103/PhysRevC.102.045801}

\bibitem[{Lopez {et~al.}(2021)Lopez, Dorso, \& Frank}]{Lopez:2020zne}
Lopez, J.~A., Dorso, C.~O., \& Frank, G.~A. 2021, Front. Phys. (Beijing), 16, 24301, \dodoi{10.1007/s11467-020-1004-2}

\bibitem[{Maggiore {et~al.}(2020)}]{Maggiore:2019uih}
Maggiore, M., {et~al.} 2020, JCAP, 03, 050, \dodoi{10.1088/1475-7516/2020/03/050}

\bibitem[{Mahmoodifar \& Strohmayer(2013)}]{Mahmoodifar:2013quw}
Mahmoodifar, S., \& Strohmayer, T. 2013, ApJ, 773, 140, \dodoi{10.1088/0004-637X/773/2/140}

\bibitem[{Melatos \& Payne(2005)}]{Melatos:2005ez}
Melatos, A., \& Payne, D. J.~B. 2005, ApJ, 623, 1044, \dodoi{10.1086/428600}

\bibitem[{Miller {et~al.}(2003)Miller, Wijnands, Mendez, Kendziorra, Tiengo, van~der Klis, Chakrabarty, Gaensler, \& Lewin}]{Miller:2002fh}
Miller, J.~M., Wijnands, R., Mendez, M., {et~al.} 2003, ApJL, 583, L99, \dodoi{10.1086/368105}

\bibitem[{Morales \& Horowitz(2024)}]{Morales:2023euv}
Morales, J.~A., \& Horowitz, C.~J. 2024, Phys. Rev. D, 110, 044016, \dodoi{10.1103/PhysRevD.110.044016}

\bibitem[{{Morales} \& {Horowitz}(2024)}]{2024arXiv240914482M}
{Morales}, J.~A., \& {Horowitz}, C.~J. 2024, arXiv e-prints, arXiv:2409.14482, \dodoi{10.48550/arXiv.2409.14482}

\bibitem[{Nandi \& Schramm(2018)}]{Nandi:2017aqq}
Nandi, R., \& Schramm, S. 2018, ApJ, 852, 135, \dodoi{10.3847/1538-4357/aa9f12}

\bibitem[{Narita {et~al.}(2001)Narita, Grindlay, \& Barret}]{Narita:2000zn}
Narita, T., Grindlay, J.~E., \& Barret, D. 2001, ApJ, 547, 420, \dodoi{10.1086/318326}

\bibitem[{Ootes {et~al.}(2019)Ootes, Wijnands, \& Page}]{Ootes:2019uvd}
Ootes, L.~S., Wijnands, R., \& Page, D. 2019, A\&A, 630, A95, \dodoi{10.1051/0004-6361/201936035}

\bibitem[{Osborne \& Jones(2020)}]{Osborne:2019iph}
Osborne, E.~L., \& Jones, D.~I. 2020, MNRAS, 494, 2839, \dodoi{10.1093/mnras/staa858}

\bibitem[{{Papaloizou} \& {Pringle}(1978)}]{1978MNRAS.184..501P}
{Papaloizou}, J., \& {Pringle}, J.~E. 1978, MNRAS, 184, 501, \dodoi{10.1093/mnras/184.3.501}

\bibitem[{Papitto {et~al.}(2013)}]{Papitto:2013hza}
Papitto, A., {et~al.} 2013, Nature, 501, 517, \dodoi{10.1038/nature12470}

\bibitem[{Patruno(2010)}]{Patruno:2010qz}
Patruno, A. 2010, ApJ, 722, 909, \dodoi{10.1088/0004-637X/722/1/909}

\bibitem[{Patruno {et~al.}(2009)Patruno, Altamirano, Hessels, Casella, Wijnands, \& van~der Klis}]{Patruno:2008km}
Patruno, A., Altamirano, D., Hessels, J. W.~T., {et~al.} 2009, ApJ, 690, 1856, \dodoi{10.1088/0004-637X/690/2/1856}

\bibitem[{Patruno {et~al.}(2017)Patruno, Haskell, \& Andersson}]{Patruno:2017oum}
Patruno, A., Haskell, B., \& Andersson, N. 2017, ApJ, 850, 106, \dodoi{10.3847/1538-4357/aa927a}

\bibitem[{Pethick {et~al.}(2020)Pethick, Zhang, \& Kobyakov}]{Pethick:2020aey}
Pethick, C.~J., Zhang, Z., \& Kobyakov, D.~N. 2020, Phys. Rev. C, 101, 055802, \dodoi{10.1103/PhysRevC.101.055802}

\bibitem[{Pi {et~al.}(2015)Pi, Yang, \& Zheng}]{Pi:2014rdd}
Pi, C.-M., Yang, S.-H., \& Zheng, X.-P. 2015, RAA, 15, 871, \dodoi{10.1088/1674-4527/15/6/009}

\bibitem[{Piraino {et~al.}(1999)Piraino, Santangelo, Ford, \& Kaaret}]{Piraino:1999qp}
Piraino, S., Santangelo, A., Ford, E.~C., \& Kaaret, P. 1999, A\&A, 349, L77.
\newblock \doarXiv{astro-ph/9910115}

\bibitem[{Pons {et~al.}(2013)Pons, Vigano', \& Rea}]{Pons:2013nea}
Pons, J.~A., Vigano', D., \& Rea, N. 2013, Nature Phys., 9, 431, \dodoi{10.1038/nphys2640}

\bibitem[{Potekhin {et~al.}(2015)Potekhin, Pons, \& Page}]{Potekhin:2015qsa}
Potekhin, A.~Y., Pons, J.~A., \& Page, D. 2015, Space Sci. Rev., 191, 239, \dodoi{10.1007/s11214-015-0180-9}

\bibitem[{Punturo {et~al.}(2010)}]{Punturo:2010zz}
Punturo, M., {et~al.} 2010, Class. Quant. Grav., 27, 194002, \dodoi{10.1088/0264-9381/27/19/194002}

\bibitem[{Reed {et~al.}(2021)Reed, Deibel, \& Horowitz}]{Reed:2021scb}
Reed, B.~T., Deibel, A., \& Horowitz, C.~J. 2021, ApJ, 921, 89, \dodoi{10.3847/1538-4357/ac1c04}

\bibitem[{Riles(2023)}]{Riles:2022wwz}
Riles, K. 2023, Living Rev. Rel., 26, 3, \dodoi{10.1007/s41114-023-00044-3}

\bibitem[{{Ruderman}(1969)}]{1969Natur.223..597R}
{Ruderman}, M. 1969, nat, 223, 597, \dodoi{10.1038/223597b0}

\bibitem[{Schatz {et~al.}(1999)Schatz, Bildsten, Cumming, \& Wiescher}]{Schatz:1999kx}
Schatz, H., Bildsten, L., Cumming, A., \& Wiescher, M. 1999, ApJ, 524, 1014, \dodoi{10.1086/307837}

\bibitem[{Schneider {et~al.}(2016)Schneider, Berry, Caplan, Horowitz, \& Lin}]{Schneider:2016zyx}
Schneider, A.~S., Berry, D.~K., Caplan, M.~E., Horowitz, C.~J., \& Lin, Z. 2016, Phys. Rev. C, 93, 065806, \dodoi{10.1103/PhysRevC.93.065806}

\bibitem[{Shapiro \& Teukolsky(1983)}]{Shapiro:1983du}
Shapiro, S.~L., \& Teukolsky, S.~A. 1983, {Black holes, white dwarfs, and neutron stars: The physics of compact objects}, \dodoi{10.1002/9783527617661}

\bibitem[{Sotani(2011)}]{Sotani:2011nn}
Sotani, H. 2011, MNRAS, 417, L70, \dodoi{10.1111/j.1745-3933.2011.01122.x}

\bibitem[{Steltner {et~al.}(2023)Steltner, Papa, Eggenstein, Prix, Bensch, Allen, \& Machenschalk}]{Steltner:2023cfk}
Steltner, B., Papa, M.~A., Eggenstein, H.~B., {et~al.} 2023, ApJ, 952, 55, \dodoi{10.3847/1538-4357/acdad4}

\bibitem[{Thompson \& Duncan(2001)}]{Thompson:2001ie}
Thompson, C., \& Duncan, R.~C. 2001, ApJ, 561, 980, \dodoi{10.1086/323256}

\bibitem[{Tripathee \& Riles(2024)}]{Tripathee:2023muh}
Tripathee, A., \& Riles, K. 2024, Phys. Rev. D, 109, 043049, \dodoi{10.1103/PhysRevD.109.043049}

\bibitem[{Tsang {et~al.}(2012)Tsang, Read, Hinderer, Piro, \& Bondarescu}]{Tsang:2011ad}
Tsang, D., Read, J.~S., Hinderer, T., Piro, A.~L., \& Bondarescu, R. 2012, Phys. Rev. Lett., 108, 011102, \dodoi{10.1103/PhysRevLett.108.011102}

\bibitem[{Ushomirsky {et~al.}(2000)Ushomirsky, Cutler, \& Bildsten}]{Ushomirsky:2000ax}
Ushomirsky, G., Cutler, C., \& Bildsten, L. 2000, MNRAS, 319, 902, \dodoi{10.1046/j.1365-8711.2000.03938.x}

\bibitem[{Van Den~Broeck(2005)}]{VanDenBroeck:2004wj}
Van Den~Broeck, C. 2005, Class. Quant. Grav., 22, 1825, \dodoi{10.1088/0264-9381/22/9/022}

\bibitem[{{Wagoner}(1984)}]{1984ApJ...278..345W}
{Wagoner}, R.~V. 1984, ApJ, 278, 345, \dodoi{10.1086/161798}

\bibitem[{Watts {et~al.}(2008)Watts, Krishnan, Bildsten, \& Schutz}]{Watts:2008qw}
Watts, A., Krishnan, B., Bildsten, L., \& Schutz, B.~F. 2008, MNRAS, 389, 839, \dodoi{10.1111/j.1365-2966.2008.13594.x}

\bibitem[{Wette(2023)}]{Wette:2023dom}
Wette, K. 2023, Astropart. Phys., 153, 102880, \dodoi{10.1016/j.astropartphys.2023.102880}

\bibitem[{{White} \& {Zhang}(1997)}]{1997ApJ...490L..87W}
{White}, N.~E., \& {Zhang}, W. 1997, ApJL, 490, L87, \dodoi{10.1086/311018}

\bibitem[{Xia {et~al.}(2023)Xia, Maruyama, Yasutake, Tatsumi, \& Zhang}]{Xia:2022rhx}
Xia, C.-J., Maruyama, T., Yasutake, N., Tatsumi, T., \& Zhang, Y.-X. 2023, Phys. Lett. B, 839, 137769, \dodoi{10.1016/j.physletb.2023.137769}

\bibitem[{Yakovlev {et~al.}(2001)Yakovlev, Kaminker, Gnedin, \& Haensel}]{Yakovlev:2000jp}
Yakovlev, D.~G., Kaminker, A.~D., Gnedin, O.~Y., \& Haensel, P. 2001, Phys. Rept., 354, 1, \dodoi{10.1016/S0370-1573(00)00131-9}

\bibitem[{{Yakovlev} \& {Urpin}(1980)}]{1980SvA....24..303Y}
{Yakovlev}, D.~G., \& {Urpin}, V.~A. 1980, \sovast, 24, 303

\bibitem[{Yim {et~al.}(2023)Yim, Gao, Kang, Shao, \& Xu}]{Yim:2023nda}
Yim, G., Gao, Y., Kang, Y., Shao, L., \& Xu, R. 2023, MNRAS, 527, 2379, \dodoi{10.1093/mnras/stad3337}

\bibitem[{Yim {et~al.}(2024)Yim, Shao, \& Xu}]{Yim:2024eaj}
Yim, G., Shao, L., \& Xu, R. 2024, MNRAS, 532, 3893, \dodoi{10.1093/mnras/stae1659}

\bibitem[{{Zhang} {et~al.}(2024){Zhang}, {Yuan}, {Wang}, \& {Ho}}]{2024MNRAS.532.4550Z}
{Zhang}, S.-R., {Yuan}, Y.-F., {Wang}, J.-M., \& {Ho}, L.~C. 2024, MNRAS, 532, 4550, \dodoi{10.1093/mnras/stae1780}

\end{thebibliography}
\bibliographystyle{aasjournal}



\end{document}